\documentclass[sigconf,screen]{acmart}
\usepackage{subfig}
\usepackage{multirow}
\usepackage{cuted}
\usepackage{capt-of}
\AtBeginDocument{%
  }

\setcopyright{acmlicensed}
\copyrightyear{2018}
\acmYear{2018}
\acmDOI{XXXXXXX.XXXXXXX}
\acmConference[Conference acronym 'XX]{Make sure to enter the correct
  conference title from your rights confirmation email}{June 03--05,
  2018}{Woodstock, NY}
\acmISBN{978-1-4503-XXXX-X/2018/06}

\begin{document}

\title{Alignment and Divergence between Humans and AI in Interpersonal Privacy Decisions}

\author{Hanxiang Zeng}
\email{hxz63@uw.edu}
\affiliation{
    \institution{University of Washington}
    \city{Seattle}
    \state{Washington}
    \country{United States}
}
\author{Shuning Zhang}
\email{zsn23@mails.tsinghua.edu.cn}
\affiliation{
    \institution{Tsinghua University}
    \city{Beijing}
    \country{China}
}
\author{Xinyuan Zhou}
\email{202311260012@mail.bnu.edu.cn}
\affiliation{
    \institution{Beijing Normal University}
    \city{Beijing}
    \country{China}
}
\author{Tianqi Song}
\affiliation{
    \institution{National University of Singapore}
    \city{Singapore}
    \country{Singapore}
}
\author{Yuhan Yuan}
\affiliation{
    \institution{Independent Researcher}
    \city{Chengdu}
    \country{China}
}
\author{Yuting Yang}
\affiliation{
    \institution{Independent Researcher}
    \city{Shanghai}
    \country{China}
}
\author{Shuai Ma}
\email{mashuai@iscas.ac.cn}
\affiliation{
  \institution{Institute of Software, Chinese Academy of Sciences}
  \city{Beijing}
  \country{China}
}
\author{Xin Yi}
\authornote{Corresponding author.}
\email{yixin@tsinghua.edu.cn}
\affiliation{
  \institution{Institute for Network Sciences and Cyberspace, Tsinghua University}
  \city{Beijing}
  \country{China}
}

\begin{abstract}
 AI assistants increasingly mediate interpersonal communication on behalf of their primary user, but they risk violating the privacy expectations of third-party information owners. Resolving these tensions requires understanding how humans anticipate interpersonal privacy boundaries. Therefore, we conducted a dyadic study (N=76) and a matched evaluation of AI models across 18 information types and 3 recipient relationships. We found that data owners' privacy judgments are highly contextual and relationship dependent. While familiar data co-owners show meaningful alignment with owners' expectations, they significantly overestimate the need for permission. Interestingly, greater familiarity within the owner–co-owner dyad was associated with both higher disclosure acceptability and lower co-owner misalignment, whereas our exploratory four-item empathy measure was not. In contrast, AI models significantly underperform human co-owners in anticipating the data acceptability, even when provided with within-dyad examples. These findings underscore a core HCI design challenge to develop privacy-aware AI that respects multi-stakeholder information boundaries.
\end{abstract}

\ccsdesc[500]{Security and privacy~Human and societal aspects of security and privacy}
\ccsdesc[500]{Human-centered computing~Empirical studies in HCI}

\keywords{Interdependent privacy, Privacy norm, Contextual Integrity, Communication Privacy Management}

\begin{teaserfigure}
 \includegraphics[width=\columnwidth]{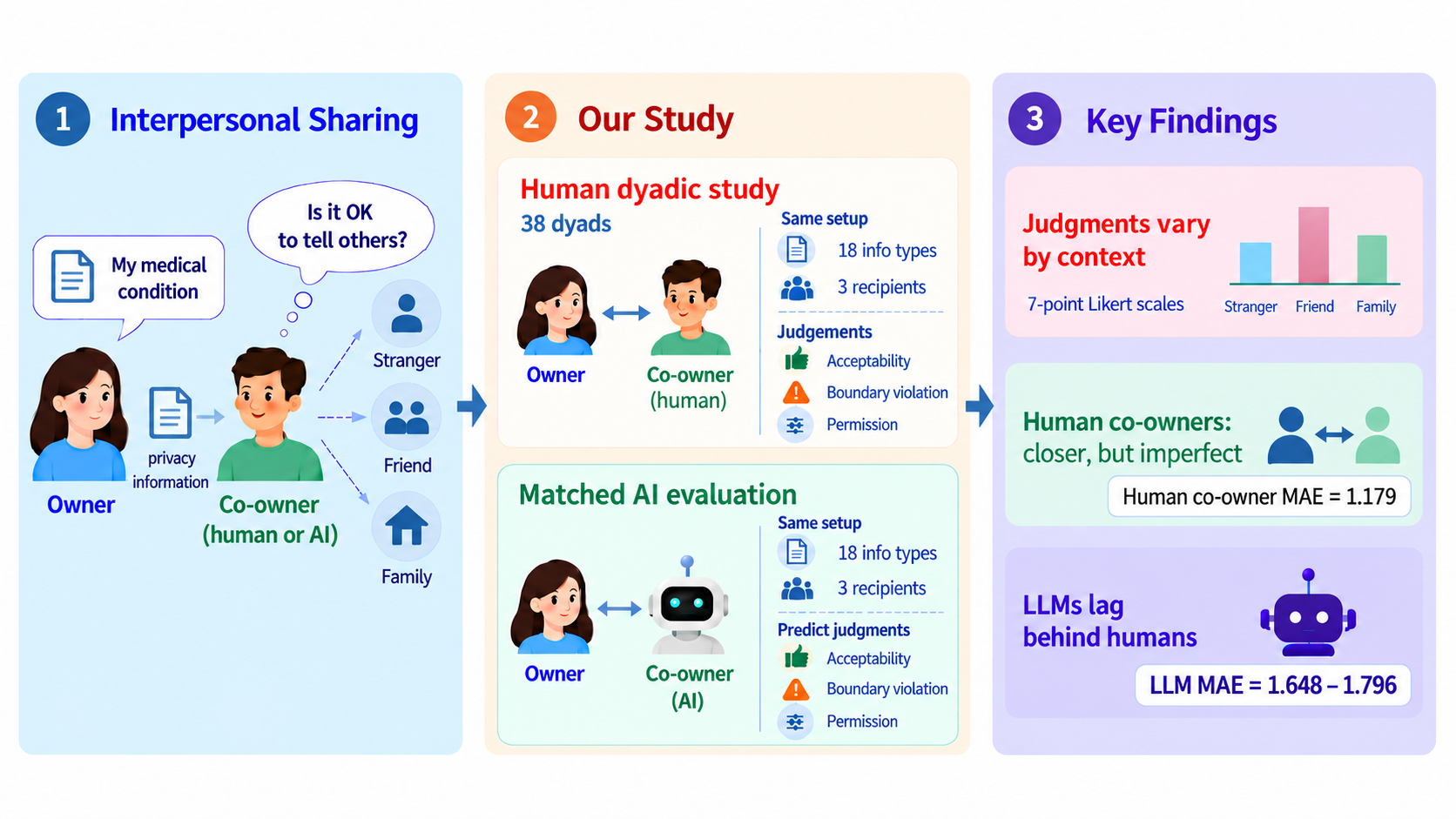}
  \caption{Illustration of our study and core findings. The image was refined by authors and GPT-5.6-sol.}
  \Description{A three-panel study overview graphic flowing left to right. Panel 1, Interpersonal sharing, shows an information owner and a co-owner (human or AI) with shared information, and icons for possible recipients: stranger, friend, and family, with the question whether it is OK to tell others. Panel 2, Our study, summarizes a human dyadic study of 38 dyads (19 friend, 19 romantic) covering 18 information types and 3 recipients, plus a matched AI evaluation where models predict acceptability, boundary violation, and need for permission. Panel 3, Key findings, highlights that judgments vary by recipient, human co-owners achieve about MAE 1.179, and LLMs remain higher-error around MAE 1.648 to 1.796.}
  \label{fig:teaser}
\end{teaserfigure}


\maketitle

\section{Introduction}

As intelligent personal assistants, such as Apple Intelligence~\cite{gunter2024apple} and Huawei Celia~\cite{huawei_celia} increasingly mediate interpersonal communication and manage communication workflows on behalf of users, privacy boundary regulation is shifting from human negotiation to AI-delegated decision making. Operating as agentic intermediaries, these systems often process and transmit information that impacts third parties. Delegating such decisions to AI introduces privacy risks that these agents violate the privacy expectations of third-party information owners, thereby compromising multi-stakeholder privacy. While recent research has begun to elicit user privacy boundaries in automated delegation~\cite{guo2025not} and benchmark interdependent privacy risks~\cite{hussain2026idpbench}, current literature primarily focuses on primary user control, leaving how AI delegation aligns with third-party privacy expectations underexplored.

Addressing this challenge requires understanding how humans themselves manage interpersonal privacy boundaries. Privacy is inherently relational, governing the flow of information among individuals~\cite{nissenbaum2004privacy}. Beyond safeguarding their own sensitive data~\cite{acquisti2015privacy}, individuals often evaluate whether it is appropriate to share others' information with third parties in collaborative or social settings, such as group chats and professional workspaces~\cite{lampinen2011together,jia2016autonomous}. Prior literature conceptualizes this phenomenon as \textit{interdependent privacy}, where an individual's privacy is subject to disclosure decisions made by others~\cite{humbert2019survey}. Grounded in CPM theory, once information is shared with a data co-owner, it becomes co-owned, requiring mutual boundary coordination and rules for subsequent disclosure~\cite{petronio2002boundaries}. However, prior work rarely examines the empirical alignment of privacy expectations between owners and co-owners, nor evaluates whether AI models can navigate these nuanced interpersonal boundaries. To bridge these empirical and algorithmic gaps, we formulate three research questions (RQs):

$\bullet$ RQ1: How do contextual factors affect data owners' acceptance of third-party disclosures made by their data co-owners?

$\bullet$ RQ2: How closely do data co-owners' privacy judgments align with information owners' judgments, and what factors are associated with this alignment?

$\bullet$ RQ3: How accurately can AI models predict information owners' privacy judgments in corresponding disclosure scenarios, relative to human co-owners?

To answer these questions, we conducted a dyadic study with 19 friend pairs and 19 romantic partner pairs, evaluating 18 information types and 3 recipient relationships. We further benchmarked 3 machine learning models, 2 prompt settings, and 7 fine-tuned models across different brands and sizes. 

For RQ1, we found that owners' privacy judgments are highly contextual and relationship-dependent. Information type significantly affected disclosure acceptability, and disclosures to friends and family were substantially more acceptable than those to strangers. 

For RQ2, familiar human co-owners showed meaningful but imperfect alignment with information owners. On a 7-point Likert scale, co-owners achieved a mean absolute error (MAE) of approximately 1.179 points across the three privacy dimensions, with about 70.7\% of judgments falling within one point of the owner's actual response. Moreover, human co-owners adopted an overly cautious attitude, overestimating the need for explicit consent. Interestingly, greater dyad-mean familiarity with the owner was associated with lower misalignment in co-owners' acceptability judgments, whereas information-specific familiarity and our brief EQ assessment were not significantly associated with misalignment.

For RQ3, our matched evaluation revealed that current AI models lack the awareness of data co-owners. Zero-shot Large Language Models (LLMs) significantly underperformed familiar data co-owners (MAE of 1.648--1.796 compared to the human baseline of 1.179), particularly struggling to navigate disclosures involving nuanced social relationships. One-shot prompting reduced average LLM MAE from 1.749 to 1.621, but all models remained significantly less accurate than human co-owners after Holm correction. Thus, a single within-dyad example was insufficient to close the human--AI gap. Furthermore, ML and SFT models generalized poorly to unseen individuals, and providing LLMs with owner-specific examples yielded inconsistent improvements that failed to close the human-AI gap. These algorithmic limitations reveal that relying on historical sharing decisions is insufficient for effective interpersonal mediation. Developing privacy-aware AI assistants requires managing interdependent privacy preferences.

Collectively, the contributions of this paper are threefold:

$\bullet$ We advance the quantitative modeling of interdependent privacy within HCI, showing that privacy boundaries shift across informational and relational dynamics.

$\bullet$ We augment CPM theory by quantitatively showing boundary anticipation misalignments between data owners and co-owners, providing a measurable framework for resolving interpersonal privacy tensions. 

$\bullet$ We advance the HCI design of privacy-aware AI by expanding the alignment focus from primary user to multi-stakeholder contexts, providing guidelines for building agents that navigate interdependent privacy boundaries. 

\section{Related Work}

We first review theoretical and empirical works on interpersonal and interdependent privacy. We then synthesize prior work on computational privacy modeling, personalized privacy assistants (PPAs), and LLM alignment.

\subsection{Interpersonal and Interdependent Privacy}

Unlike individual privacy management, privacy in networked environments is continually negotiated across social relationships. Within this social perspective, researchers study privacy through two related lenses: interpersonal privacy and interdependent privacy. \textbf{Interpersonal privacy refers to how individuals manage self-disclosure and establish behavioral boundaries directly with others in social relationships~\cite{lampinen2011together}. Interdependent privacy refers to situations where one person's data sharing decision directly affects another person's privacy~\cite{humbert2019survey,data2013interdependent}.} The key difference is that interpersonal privacy centers on direct social norms and mutual expectations between interacting individuals, while interdependent privacy focuses on the external privacy consequences that one person's action imposes on someone else. \textbf{We primarily focus on interdependent privacy, as our work examines how a data co-owner's disclosure to third parties affects the information owner's privacy boundaries.} 

In multi-stakeholder contexts, interdependent privacy is closely related to, yet conceptually distinct from, \textit{bystander privacy}. Bystander privacy concerns third parties whose personal data are passively captured by another user's device or sensing system~\cite{o2023privacy,marky2020don}. Prior work has extensively investigated bystanders' privacy requirements~\cite{marky2020don}, highlighted the operational challenges of ambient sensing~\cite{o2023privacy}, synthesized emerging technological landscapes~\cite{saqib2025bystander,yao2019privacy}, and proposed tangible or interface-level mitigations~\cite{ahmad2020tangible}. However, while bystander privacy centers on passive exposure and asymmetrical awareness between strangers or observers, interdependent privacy arises from relational co-ownership, where data co-owner evaluates when and how to disclose shared information to third parties.

To understand how interpersonal privacy boundaries are negotiated, prior work conceptualizes this dynamic as boundary regulation~\cite{palen2003unpacking}. Extending this view, CPM theory posits that disclosing private information establishes co-ownership, requiring parties to coordinate rules for subsequent access and redistribution~\cite{petronio2002boundaries}. Empirical research on social network services shows that individuals employ both individual and collaborative strategies to govern disclosures involving themselves and others~\cite{lampinen2011together}. Consequently, receiving personal data carries expectations regarding its future circulation~\cite{palen2003unpacking,petronio2002boundaries,lampinen2011together}. Sannon et al.~\cite{sannon2020just} extended CPM to conversational systems, showing that users perceived agents more negatively when response data was shared with advertisers and third parties. Zhang et al.~\cite{zhang2026privacy} also investigated data sharing with smart home agents, quantifying data transmission and sharing boundaries. 

However, when information boundaries intersect, individual disclosure decisions can generate privacy externalities. Interdependent and group privacy research shows that one person's disclosure, authorization, or sharing action directly alters the privacy posture of others~\cite{choksi2024groups,humbert2019survey}. Resolving these multi-party conflicts requires negotiating divergent preferences across shared content~\cite{jia2016autonomous,such2017photo}, often by integrating contextual parameters, multi-stakeholder priorities, and explicit user rationale into collective sharing policies~\cite{fogues2017sharing}. Stakeholder misalignments also emerge across asymmetrical technological interfaces. For example, bystander expectations surrounding smart glasses diverge markedly from wearers' transparency practices~\cite{wang2026mind,zhang2026visguardian}. While these works highlight privacy conflicts among multiple stakeholders, they do not directly examine whether data owners' and co-owners' privacy expectations align.

Contextual Integrity (CI) theory formalizes this relational complexity by defining privacy as the appropriateness of information flow along five core parameters: data subject, sender, recipient, information attribute, and transmission principle~\cite{nissenbaum2004privacy,tran2025understanding}. Researchers have broadly leveraged CI to unpack boundaries on social platforms~\cite{shi2013contextual}, showing that perceived sharing appropriateness depends heavily on recipient relationships and relational closeness~\cite{wiese2011close}. Crucially, generalized social norms may differ from an individual's personal privacy preferences~\cite{hoyle2020norms}; a disclosure deemed broadly acceptable by group standards may still conflict with an individual owner's preferred boundary. Furthermore, research on interpersonal perception indicates that judgments about a close partner can reflect both genuine accuracy and systematic bias~\cite{kenny2001accuracy}. This gap motivates investigating the alignment between co-owner's privacy judgments and data owner's expectations in direct social sharing contexts.

\subsection{Computational Privacy Modeling, Assistance, and LLM Alignment}
Early computational privacy research focused on mitigating decision fatigue through recommendation systems, behavioral nudges, and adaptive permission profiles~\cite{ayci2022uncertainty,fang2010privacy,liu2016follow,wang2013privacy}. PPAs built upon this foundation by encoding user preferences and automating privacy enforcement across systems with varying degrees of autonomy~\cite{marky2024decide,morel2025ai}. However, deploying PPAs across diverse populations presents structural hurdles: user acceptance varies by technological affinity and desired agency~\cite{stover2023investigating}, and cross-cultural evaluations indicate that PPA preference constructs exhibit measurement and compositional variance across regions~\cite{xu2026acceptance}. Recent efforts have consequently targeted human-AI bidirectional alignment to better capture evolving user intent~\cite{zhang2025towards,zhang2025situguard,zhang2025evaluating}.

With the emergence of LLMs, privacy evaluation shifted from static access control to natural language reasoning and leakage detection. Extensive benchmarking demonstrates that standard LLMs and autonomous agents persistently leak sensitive details in both single-turn and complex multi-turn interactions~\cite{juneja2025magpie,li2024privlm,mukhopadhyay2025privacybench}. Privacy-preserving prompts and localized prompt-reformulation frameworks mitigate, but do not eliminate, this leakage~\cite{ngong2025protecting,mukhopadhyay2025privacybench}. Moreover, multi-agent orchestration can restrict unprompted exposure~\cite{li20251}, yet underlying privacy representations remain fragile and easily degrade under task-specific fine-tuning~\cite{goel2026privacy}.

To evaluate semantic privacy reasoning, recent benchmarks formalize contextual norms via legal statutes and CI criteria~\cite{cheng2024cibench,fan2024goldcoin,li2025privaci,li2025privacy}. Although state-of-the-art models identify basic CI parameters, they struggle with multi-topic contexts and frequently violate contextual norms in situations where human users maintain strict boundaries~\cite{cheng2024cibench,mireshghallah2024can,shao2024privacylens}. These are further complicated by the models' inferential capabilities~\cite{zhang2026pervasive,zhang2025through}. Moreover, in multi-party and collaborative environments, where benchmarks such as MuPPET and MAGPIE indicate that information leakage escalates significantly when conversational dynamics involve multiple interlocutors~\cite{juneja2025magpie,ruzzetti2026muppet}.

Beyond normative evaluations, an emerging branch of research explores person-specific simulation. While personalizing LLMs via demographic attributes and explicit privacy attitudes enhances simulation fidelity, models still struggle to reproduce authentic individual behaviors~\cite{flemings2026privacysim}. Although privacy benchmarks such as IDP-Bench examine LLM reasoning over co-owned data~\cite{hussain2026idpbench}, existing frameworks rarely address whether an AI can accurately anticipate an information owner's subjective approval of onward personal disclosures made by a familiar peer.


\section{Investigating Interdependent Privacy Awareness of Users and AIs}

We conducted a dyadic scenario-based survey and corresponding modeling using AI to investigate interdependent privacy awareness.

\subsection{Survey Design}

To examine interdependent privacy judgment, we designed a dyadic scenario-based survey. Within each dyad, the two participants alternated between data owner and data co-owner. The \textit{data owner} evaluated an information sharing scenario from their own perspective, while the paired \textit{data co-owner} responded to the same scenario. The roles were subsequently reversed so that both participants served as the owner and co-owner.

After human data collection, we applied the same task to AI systems, where AI acts as data co-owners, simulating the human data co-owner's personal secretary. The AI models received the corresponding contextual information and were asked to predict either data owner's responses, or the data co-owner's misalignment level against the data owner. This matched design enables comparisons among owner judgments, co-owner judgments, and the modeling using AI.

\subsubsection{Contextual Scenario Design}

Our scenario design was grounded in CI, which characterizes privacy in terms of contextual information flows involving an information subject, sender, recipient, information attribute, and transmission conditions~\cite{nissenbaum2004privacy}. Therefore, we constructed scenarios that varied the information type and third-party recipient. Besides, each vignette held the sender (the co-owner), the information subject (the owner), the channel (a message), and the transmission principle (permission not obtained) constant.

\begin{table*}[!htbp]
  \centering
  \small
  \caption{Eighteen information types used in our study, grouped into three higher-level categories.}
  \label{tab:information_types}
  \begin{tabular}{p{0.29\linewidth} p{0.26\linewidth} p{0.4\linewidth}}
    \toprule
    \textbf{Identity \& Contextual Information} &
    \textbf{State \& Behavioral Information} &
    \textbf{Cognitive \& Intimate Information} \\
    \midrule
    Name & Financial information & Beliefs/thoughts \\
    Identification information & Health information & Family/relationships \\
    Online identity & Preferences/interests & Shared private moments with the co-participant \\
    Affiliation & Behavioral habits & Shared private moments with a friend \\
    Email address &  & Shared private moments with a romantic partner \\
    Phone number &  &  \\
    Address &  &  \\
    Geolocation &  &  \\
    Time information &  &  \\
    \bottomrule
  \end{tabular}
\end{table*}

A co-owner therefore disclosed a specified type of the owner's information to one recipient without asking permission.
These fixed parameters align with existing designs in privacy research~\cite{martin2016measuring,shvartzshnaider2024privacy}. First, we selected ``a message'' as text-based messaging is the prevalent medium for everyday interpersonal data sharing and is highly susceptible to privacy leaks~\cite{lampinen2011together}. Second, fixing the transmission principle to an unauthorized disclosure is important for inducing boundary turbulence. Grounded in CPM~\cite{petronio2002boundaries}, presenting a scenario where permission is bypassed allows us to quantify the information owner's expectations for consent, and evaluate the severity of the resulting boundary violation.

For information types, we used a common personal data taxonomy as the base~\cite{zhou2025rescriber}, and augmented it with lifestyle-related information, which may be sensitive to privacy, but may not be directly identified~\cite{chua2021effects}. We further augmented three types of relational data which are not specifically held by one data owner, but may be private. We categorized these 18 privacy information to three dimensions for later analysis:\textit{identity and contextual information}, \textit{state and behavioral information}, and \textit{cognitive and intimate information}. Detailed information types are shown in Table~\ref{tab:information_types}.

\paragraph{Identity and contextual information.}
This category contains information that identifies, locates, contacts, or situates an individual within social or institutional. It includes \textit{name}, \textit{identification information}, \textit{online identity}, \textit{affiliation}, \textit{email address}, \textit{phone number}, \textit{address}, \textit{geolocation}, and \textit{time information}. Identification information refers to formal identifiers that can distinguish or verify an individual. Online identity captures user names, account identities, or online profiles. Time information captures temporal information about when a person was present, active, available, or engaged in an activity.

\paragraph{State and behavioral information.}
This category contains information describing a person's current state or recurring behavioral patterns. It includes \textit{financial information}, \textit{health information}, \textit{preferences/interests}, and \textit{behavioral habits}. Preferences and interests describe personal tastes, likes, dislikes, or interests, whereas behavioral habits refer to recurring routines or activity patterns.

\paragraph{Cognitive and intimate information.}
This category contains internal states and relationally embedded information that are dependent on interpersonal context. It includes \textit{beliefs/thoughts}, \textit{family/relationships}, \textit{shared private moments with the co-participant}, \textit{shared private moments with a friend}, and \textit{shared private moments with a romantic partner}. Beliefs and thoughts capture opinions, values, or internal viewpoints, while family and relationship information concerns interpersonal or family circumstances. Shared private moments refer to experiences or information jointly associated with the owner and another person. Unlike traditional personal information that primarily concerns one individual, these categories explicitly represent information whose privacy implications may involve multiple people.

For third-party recipients, we selected family members, friends, and strangers to operationalize strong, weak and no relationships respectively~\cite{granovetter1977strength}. These representative categories align with established taxonomies in prior work~\cite{yu2018my,amirkhani2026talking,wang2026mind}. We acknowledge that other relationship types exist, such as classmates and colleagues, and defer them to future work. 
For each dyad, they rated five randomly-selected information type from the 18-type pool. Each selected information type was instantiated with three third-party recipient relationships: \textit{stranger}, \textit{friend}, and \textit{family member}. Thus, one ownership direction contained 5 information types $\times$ 3 recipients = 15 scenario evaluations. Because the two participants exchanged owner and co-owner roles, each dyad completed 2 ownership directions $\times$ 5 information types $\times$ 3 recipients = 30 scenario evaluations. Scenario presentation order was randomized within each assigned set to reduce ordering effects. We empirically selected five as the number of information type to reduce fatigue and ensure proper engagement. 
For example, when Participant A served as the information owner, Participant A reported whether they accept information sharing by Participant B, and Participant B uttered their own judgment around Participant A's data. When the roles were reversed, Participant B became the information owner and Participant A the co-owner. This structure allowed us to evaluate interdependent privacy alignment in both directions within each dyad.

To protect participants' privacy, participants could either enter it through the study interface or communicate it privately outside the platform. Information shared privately was not recorded by the system.

\begin{figure*}[!htbp]
    \centering 
    \includegraphics[width=0.8\textwidth]{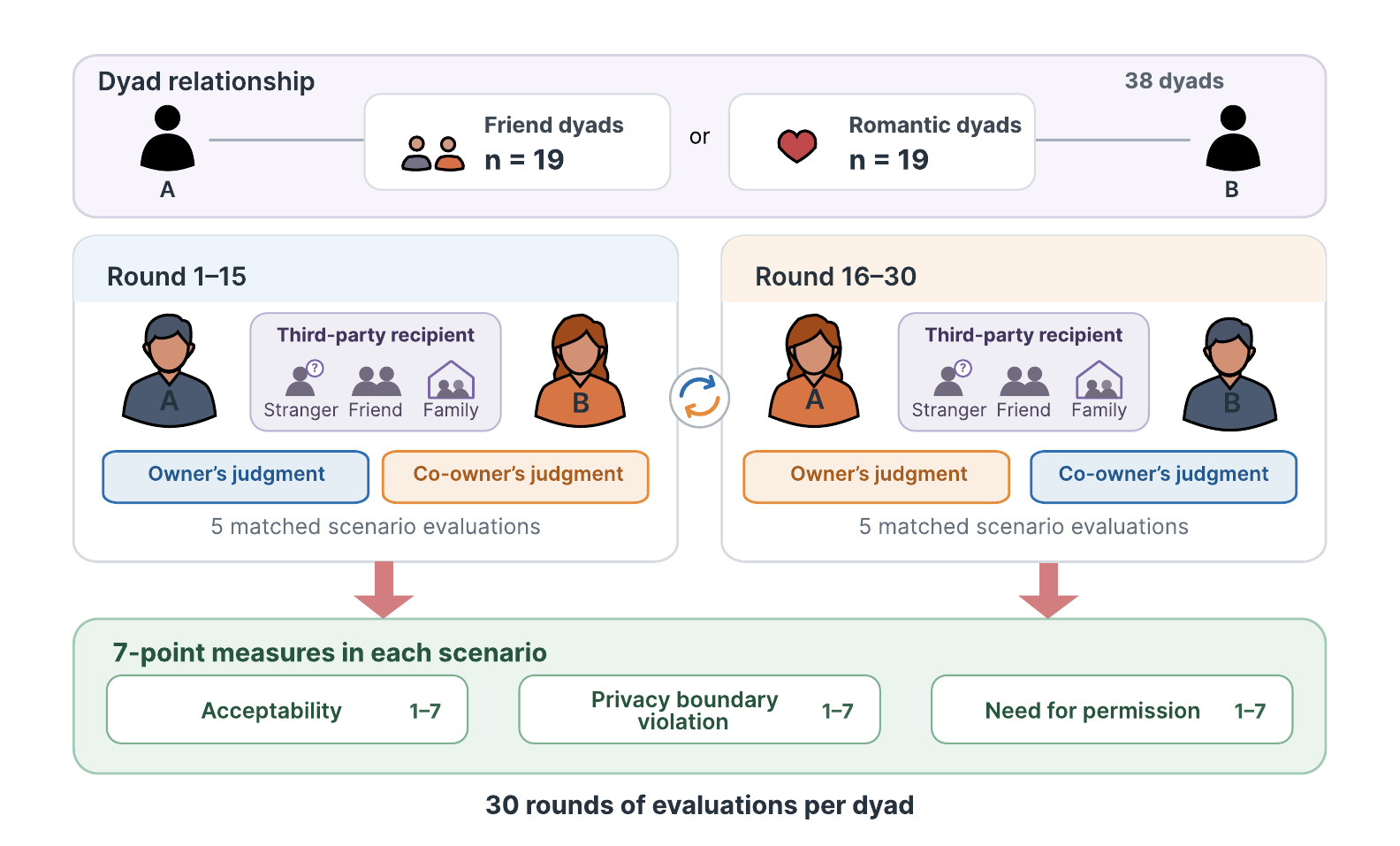}
    \caption{Illustration of the experiment procedure.}
    \Description{A flowchart of the human dyadic study procedure. At the top, 38 dyads split into 19 friend dyads and 19 romantic dyads, each with an information owner and an information co-owner. The middle shows two blocks of scenario evaluations: rounds 1 to 15 with one role assignment and rounds 16 to 30 after roles switch, for 30 evaluations per dyad. Each scenario involves a third-party recipient (stranger, friend, or family). At the bottom, both roles rate each scenario on three 7-point scales: acceptability, privacy-boundary violation, and need for permission.}
    \label{fig:procedure}
\end{figure*}

\subsubsection{Dyadic and Information Familiarity Assessment}

Before the main scenario task, we adopted a pre-study questionnaire to record whether the dyad consisted of friends or romantic partners, and collected interpersonal familiarity measures. The pre-study questionnaire additionally included empathy assessment scale. We used an exploratory four-item empathy measure (Appendix~\ref{sec:appendix_d_empathy}). The assessment was informed by constructs in the Empathy Quotient (EQ) ~\cite{baroncohen2004eq} and related perspective-taking measures~\cite{davis1983iri}, and contextualized to the dyadic privacy setting. The EQ-related rating was calculated as the mean of the four positively worded EQ items, with higher ratings indicating higher self-reported EQ. These measures were collected before the main task so that we could examine whether interpersonal characteristics or information-specific knowledge were associated with privacy judgments or alignment.

Participants also reported their familiarity with their partner's reported specific information used in the study. We distinguished this \textit{information familiarity} from general interpersonal familiarity, where a participant may know their partner well overall while having limited knowledge of a particular type of information about them.

\subsubsection{Privacy Judgments}

For each scenario, participants as information owners evaluated the co-owner's unauthorized disclosure across three dimensions using 7-point Likert scales. Grounded in CI~\cite{nissenbaum2004privacy}, we assessed \textit{acceptability} (1=completely unacceptable, 7=completely acceptable) to evaluate context-relative appropriateness. Drawing on CPM theory~\cite{petronio2002boundaries}, we further captured boundary coordination via \textit{privacy boundary violation} (1=not at all violated, 7=strongly violated) and \textit{need for permission} (1=strongly disagree, 7=strongly agree). Across all measures, higher scores indicate stronger perceived acceptability, violation, and requirements for prior consent, respectively.

\subsection{AI Model Selection}
\label{sec:model-selection}
We evaluated both machine learning models and LLMs for modeling owners' privacy expectation and co-owners' alignment. Different settings parallel to human co-owners with different familiarity, such as making decisions based on prior histories. 

\textbf{Machine learning models setting.}
We also evaluated three traditional machine-learning models: linear regression, gradient-boosted decision trees (GBDT) and random forest. These models were used to predict the information owner's privacy judgments and acceptability gap between owner and co-owner under different train-test settings.

\textbf{Zero-shot and One-shot LLM settings.} We report three primary LLMs in detail (GPT-5.5, Qwen3-8B, and Qwen3.6-27B) and apply the same zero- and one-shot protocol to a broader set of ten models across different brands for robustness check: Gemini Pro, Claude Opus 4.8, Grok 4.6, GPT-5.5, GPT-6-Astra, GPT-5.6-Sol, GPT-5.4, GPT-5.6-Terra, Qwen3-8B, and Qwen3.6-27B.
We adopt two prompting settings. In the \textit{zero-shot} setting, the model received the task and contextual information without examples of the target dyad's responses. In the \textit{one-shot} setting, we randomly sampled one of the 30 scenarios from the target dyad, provided both the owner's and the co-owner's three privacy ratings for that scenario, and asked the model to predict the remaining scenarios. The demonstration trial was excluded from scoring; for fair comparison, zero-shot metrics paired with one-shot analyses were recomputed on the same non-demonstration scenarios (87 matched judgments per dyad). Demonstrations were always within-dyad. Because each dyad has two information owners, a demonstration may be person-matched to the target trial or come from the other partner. However, as these two cases did not differ significantly in error (dyad-level paired test on averaged MAE of models: $\Delta\approx$-0.160, t(37)=-1.84, p= .074), so we pool them in all one-shot analyses. The same protocol was applied to the 10 LLMs evaluated in Section~\ref{sec:rq3}. This comparison allowed us to examine whether limited owner-specific information can improve the alignment of the model's privacy modeling capability.

\textbf{Low-Rank Adaptation (LoRA) setting.} We fine-tuned seven models across different brands and model sizes (i.e., Qwen2.5-0.5B, Qwen2.5-1.5B, Qwen2.5-3B, Qwen2.5-7B, Mistral-7B, SmolLM2-1.7B, and Phi-3.5-mini) to predict privacy-sharing acceptability using parameter-efficient LoRA adaptation. We framed the task as a continuous rating prediction on the original 1–7 scale. Each input prompt encoded the interpersonal relationship, recipient type, information category, anonymization status, familiarity measures, EQ-related ratings, the information owner's contextual text, and the corresponding sharing scenario. The models were fine-tuned to autoregressively generate the target rating. We configured LoRA with a rank of 32, $\alpha = 64$, and a dropout of 0.05. Models were trained for one epoch with a learning rate of $5\times 10^{-5}$, weight decay of 0.01, bfloat16 precision, a per-device batch size of 1, and gradient accumulation over 8 steps, yielding an effective batch size of 8. The dataset included 1,140 instances across 38 dyads, evaluated via five-fold GroupKFold cross-validation ensuring that dyads were separated between training and test sets. Evaluation metrics included Mean Absolute Error (MAE), exact match accuracy, and within-$\pm1$ accuracy.


\subsection{Measures}




We distinguish between \textit{misalignment}, which captures the magnitude of disagreement, and \textit{signed misalignment}, which captures the direction of disagreement.

\textbf{Human misalignment.} For each privacy dimension, human misalignment was calculated relative to the corresponding owner's response. Our magnitude-based measure was the mean absolute error between the co-owner judgment and owner judgment. For directional analyses, we define signed error as the co-owner’s judgment minus the owner’s judgment. For acceptability, a negative signed error indicates that data co-owner's judgment is less accepting than the data owner's judgment. For privacy boundary violation and need for permission, a positive signed error indicates that the data co-owner's judgments express stronger privacy concern than the owner's judgment.

\textbf{AI prediction error.} AI prediction error was defined analogously: $E_{\mathrm{AI}} = \mathrm{AIPrediction} - \mathrm{OwnerJudgment}$ This measure allows us to evaluate whether AI systems over- or underestimate owner judgments and to compare the direction and magnitude of AI errors with those of human partners.

\textbf{Prediction gap between AI and data co-owner.}
We additionally compare AI predictions directly with human partner's judgments. This comparison captures whether AI systems produce different privacy judgments from people who personally know the information owner. We interpret the directional differences separately for the three judgment dimensions. 


We additionally report exact-match alignment and within-one-point alignment~\cite{wu2025personalized}. Exact match measures the proportions that co-owners' response exactly equals the owner's response~\cite{flemings2026privacysim}. Within-one-point alignment measures the proportion of responses falling within $\pm1$ point of the owner's response on the 7-point scale, capturing cases where the guess is close enough to still be useful in practice. Pearson correlations were used as a complementary measure of whether predicted judgments track which cases the owner finds more or less acceptable, even when absolute scores differ.

\subsection{Participants and Recruitment}

We recruited participants from Prolific. Eligible participants were required to participate with a friend or romantic partner, and both members of the dyad were required to have active Prolific accounts. We focused on friendships and romantic partnerships as two relationship contexts involving interpersonal privacy concerns. ~\cite{amirkhani2026talking,yu2018my}. While other social relationships (e.g., colleagues, classmates) warrant investigation, we leave them to future work. Strangers were excluded because dyadic boundary regulation typically requires an existing social relationship. Information disclosure involving strangers often causes outright privacy violations. We recruited 76 participants through Prolific, forming 38 dyads: 19 friend dyads and 19 romantic-partner dyads. Further participant demographics and recruitment materials are provided in Appendices~\ref{app:demographics} and~\ref{app:recruitment}.

Table~\ref{tab:demographics} summarizes the demographics of 76 Prolific-recruited participants in the analytic sample. Each participant was compensated 11 GBP for their time according to the recommended compensation standard of Prolific. The study was advertised as taking approximately 40 minutes. Recorded dyad completion times averaged 39.1 minutes (range: 12.4--143.9 minutes). The study was approved by our university's Institutional Review Board (IRB).

\subsection{Pilot Study}

We conducted two rounds of pilot study through convenience sampling, with each round involving 3-6 pairs of participants with different familiarities. Participants completed the study, and raised questions regarding the study design and questionnaire descriptions. We refined the questionnaire detail to make it more understandable after the pilot study. Besides, through the pilot study we determined that each group should rate 5 information types, which lasted about 20-30 minutes.

\subsection{Data Analysis}

Dyad-level comparisons used paired or Welch's t-tests. Within-dyad recipient contrasts used Friedman tests, and one-sample t-tests on dyad-mean signed bias. Complementary trial-level analyses included Kruskal-Wallis tests across information categories and multivariable ordinary least squares regressions, which incorporated information category, standardized dyad-mean interpersonal familiarity, EQ score, recipient role, and sharer type. Human regressions use classical OLS SEs; the LLM contextual analysis below uses dyad-clustered SEs. Post-hoc pairwise comparisons were adjusted using the Holm correction. Statistical significance is reported at $p< .05$ and denoted in figures as follows: * $p< .05$, ** $p< .01$, *** $p< .001$, and ns $p\geq.05$.



\section{RQ1: Information Owners' Privacy Judgments}

We examine data owners' privacy decisions progressively, progressing from overall rating distributions to contextual effects and multivariate regressions.

\subsection{Overall Ratings}

\textbf{\textit{Data owners' acceptability spanned a broad spectrum, but were strongly correlated with perceived boundary violation and permission needs.}} Overall, owners rated with a quite loose range from 1 to 7 for acceptability (trial-level $M=4.03$, $SD= 2.26$). At the scenario level, owner acceptability was negatively correlated with boundary violation (Pearson $r= -.849$, $p< .001$) and need for permission ($r= -.695$, $p< .001$). The corresponding dyad-level correlations were $r= -.872$ ($p< .001$) and $r= -.572$ ($p< .001$), respectively. Interestingly, in a few cases such as address (Acc \textit{M}= 4.12, Consent \textit{M}= 5.40; 32.1\% of trials with Acc$\geq$5 and Consent$\geq$5) and geolocation (Acc \textit{M}= 4.13, Consent \textit{M}= 5.41; 18.5\%), they not only show a high acceptability, but also a high need for consent. This indicates that even when data sharing is deemed acceptable, owners may still demand explicit notification.

\begin{figure*}[!htbp]
    \centering 
    \includegraphics[width=\textwidth]{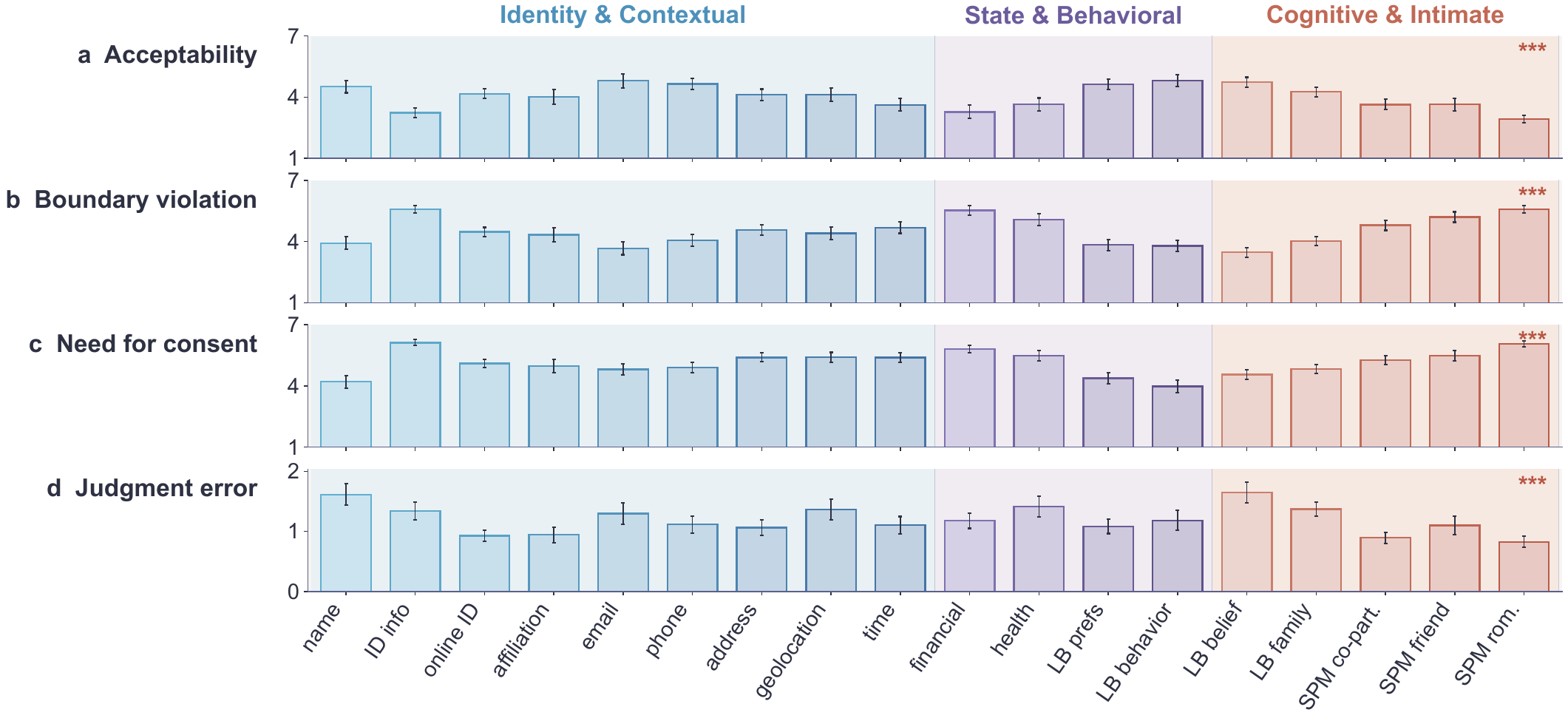}
    \caption{(a) Data acceptability, (b) boundary violation, and (c) need for consent, and (d) misalignments across data types. Asterisks denote Kruskal--Wallis tests across information types ($p< .001$). Errorbar indicated one standard error (SE).}
    \Description{Four aligned bar charts across eighteen information types grouped into Identity and Contextual, State and Behavioral, and Cognitive and Intimate. Panel (a) shows mean owner acceptability on a 1 to 7 scale; panel (b) boundary violation; panel (c) need for consent; panel (d) prediction or misalignment error on a lower scale. Cognitive and intimate types tend to look less acceptable and more sensitive. Each bar has a standard-error whisker, and significance marks appear for group differences.}
    \label{fig:transmission_recipient}
\end{figure*}

\subsection{Effects of Contextual Factors}

We analyzed effects of factors on owners' acceptability, first focusing on CI-related factors. \textbf{\textit{Information type significantly affected data owners' acceptability, with identity and behavioral contexts receiving higher acceptability than intimate disclosures.}} A Kruskal--Wallis test revealed a significant effect of information type on acceptability ($H(17) = 74.03, p < .001$). Descriptively, \textit{email address} and \textit{behavior habit} received the highest acceptability ($M= 4.81$ and $4.81$). \textit{Identification} and \textit{shared private moments with a romantic partner} data received the lowest acceptability ($M= 3.24$ and $2.93$). At the broader level, \textit{Identity \& Contextual} ratings were descriptively higher than \textit{Cognitive \& Intimate ratings}, although the difference was not significant ($t(32)= 1.92$, $p_{Holm}= .192$). \textit{Identity \& Contextual} versus \textit{State \& Behavioral} ($t(25)= 0.71$, $p_{Holm}= .485$) and \textit{State \& Behavioral} versus \textit{Cognitive \& Intimate} ($t(23)= 1.62$, $p_{Holm}= .238$) also showed no significant differences.

\textbf{\textit{Transmission recipients dictated privacy decisions, as sharing with strangers was less acceptable than with close relational ties.}} Transmission recipient showed a significant effect on owner acceptability (Friedman $\chi^{2}(2)=52.90$, $p< .001$), with friend ($M= 4.58$) and family ($M= 4.76$) rated significantly higher than stranger ($M= 2.75$; post-hoc friend vs. stranger $p_{Holm}< .001$; family vs. stranger $p_{Holm}< .001$). In contrast, acceptability ratings between friends and family members did not differ significantly ($p = .114$), underscoring a sharp difference between familiar social circles and strangers.

\textbf{\textit{Data acceptability remained consistent across partner contexts, interpersonal familiarity, information-specific familiarity and individual's EQ-like ratings.}} Specifically, acceptability ratings did not differ significantly between friend dyads ($M = 4.03$) and romantic-partner dyads ($M = 4.03$; Welch's $t(33.99) = 0.01, p = .992$). This indicates that both close relational structures exhibit comparable baseline acceptability. Simple bivariate correlations similarly failed to reach significance for individual familiarity with the data owner ($r = .189, p = .256$), familiarity with specific information ($r = .055, p = .743$), or overall EQ scores ($r = .16, p = .33$).

\subsection{Owner Acceptability Regression}

Beyond single-factor correlation, we further used multivariable regression models to assess how contextual and individual-level factors were associated with owner's acceptability. 

\begin{figure*}[!htbp]
    \centering 
    \subfloat[Owners' acceptability.]{
       \includegraphics[width= 0.5\textwidth]{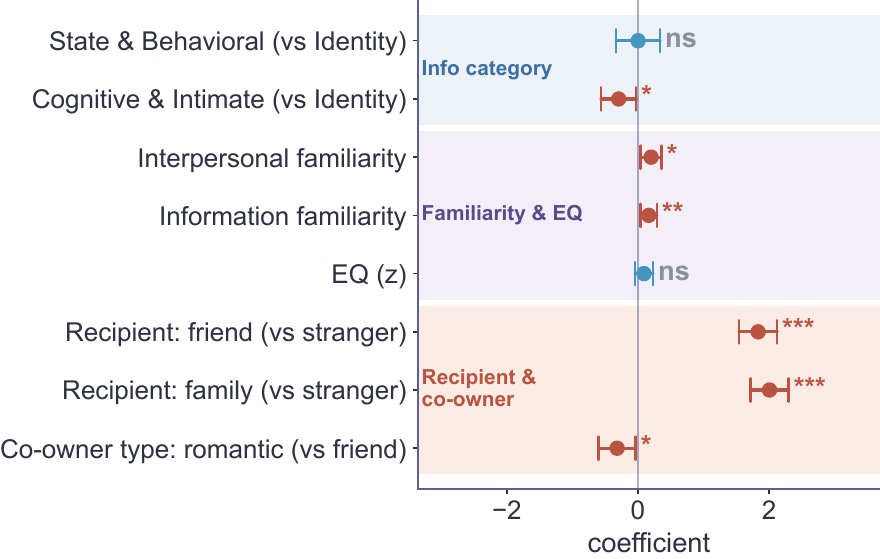}
        \label{fig:owner_prediction}
    }
    \subfloat[Co-owners' misalignments.]{
        \includegraphics[width= 0.5\textwidth]{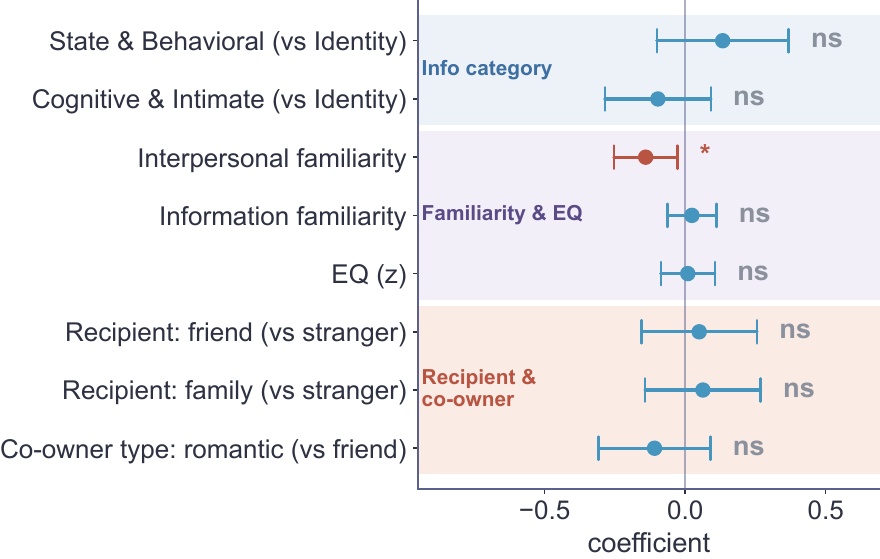}
        \label{fig:coowner_prediction}
    }
    \caption{Regression analysis of (a) owners' acceptability and (b) co-owners' misalignment across different factors. Errorbar denoted 95\% CI.}
    \Description{Two side-by-side forest plots showing regression coefficients with 95 percent confidence intervals. Panel (a) models owners' acceptability. Cognitive and intimate information is negatively associated with acceptability relative to identity and contextual information. Familiarity with the owner and information-specific familiarity are positively associated with acceptability, whereas the exploratory EQ score is not significant. Disclosures to friends and family are positively associated with acceptability relative to disclosures to strangers, while romantic-partner co-owners are negatively associated with acceptability relative to friend co-owners. Panel (b) models co-owner misalignment. Familiarity with the owner is negatively associated with misalignment, while information category, information-specific familiarity, EQ, recipient condition, and sharer type are not statistically significant.}
    \label{fig:prediction}
\end{figure*}

\textbf{Information type strongly correlated with acceptability, with cognitive and intimate disclosures significantly reducing acceptability.} As shown in Figure~\ref{fig:owner_prediction}, relative to \textit{Identity \& Contextual} data, the disclosure of \textit{Cognitive \& Intimate} information was negatively associated with owner acceptability ($\beta = -0.294$, 95\% CI [$-0.561, -0.027$], $p = .031$), whereas \textit{State \& Behavioral} data showed no significant effect ($\beta = 0.003$, 95\% CI [$-0.330, 0.336$], $p = .987$). This negative association highlights heightened boundaries when handling personal or shared thoughts.

\textbf{Interpersonal familiarity and dyadic structure significantly conditioned sharing decisions, whereas individual empathy levels exerted no effect.} Specifically, information-specific familiarity ($\beta = 0.165$, 95\% CI [$0.041, 0.289$], $p = .009$) and overall dyadic interpersonal familiarity ($\beta = 0.201$, 95\% CI [$0.041, 0.361$], $p = .014$) were positive predictors of acceptability. Dyadic relationship type was not significantly associated with LLM-predicted acceptability ($\beta = 0.042$, 95\% CI [$-0.187$, $0.270$], $p = .714$). Meanwhile, dyad-mean EQ scores failed to reach significance ($\beta = 0.092$, 95\% CI [$-0.044, 0.229$], $p = .185$). These results provide no clear evidence that LLM-predicted acceptability differed between friend and romantic partner dyads.

\textbf{Recipient type was the primary influential factor of acceptability, as sharing with strangers was less acceptable than with known social relationships.} Disclosing information to friends ($\beta = 1.834$, 95\% CI [$1.542, 2.126$], $p < .001$) or family members ($\beta = 2.008$, 95\% CI [$1.716, 2.300$], $p < .001$) was significantly more acceptable than sharing with strangers. 

\section{RQ2: Human Co-Owner Alignment}

To address RQ2, we evaluate how well data co-owners and owners' expectations align. We proceed from overall rating distributions to the effects of contextual factors.

\subsection{Overall Ratings}

\textbf{Data co-owners showed moderate but imperfect alignment with information owners, and were biased towards conservative ratings}. Across three dimensions, data co-owners showed moderate alignment with owner acceptability responses (MAE$= 1.179$, exact match$=40.4\%$, and within-one-point alignment$=70.7\%$). For acceptability, boundary violation, and need for permission, respectively, MAEs were 1.230, 1.146, and 1.161; exact-match rates were $=41.5\%$, $=41.1\%$, and $=38.7\%$; and within-one-point rates were $=69.2\%$, $=72.4\%$, and $=70.6\%$. Signed errors were descriptively conservative. Co-owners rated acceptability 0.153 points lower than owners on average, but this bias was not statistically significant ($t(37)=-1.99$, $p= .054$). Boundary-violation ratings were 0.147 points higher, also not statistically significant ($t(37)= 1.89$, $p= .066$). Co-owners rated the need for permission 0.177 points, which is significantly higher than owners ($t(37)= 2.81$, $p= .008$; Holm-adjusted $p= .024$ across the three bias comparisons). Thus, the clearest evidence of directional conservatism was the need for permission. 

\begin{figure*}[!htbp]
    \centering 
    \includegraphics[width= 0.6\textwidth]{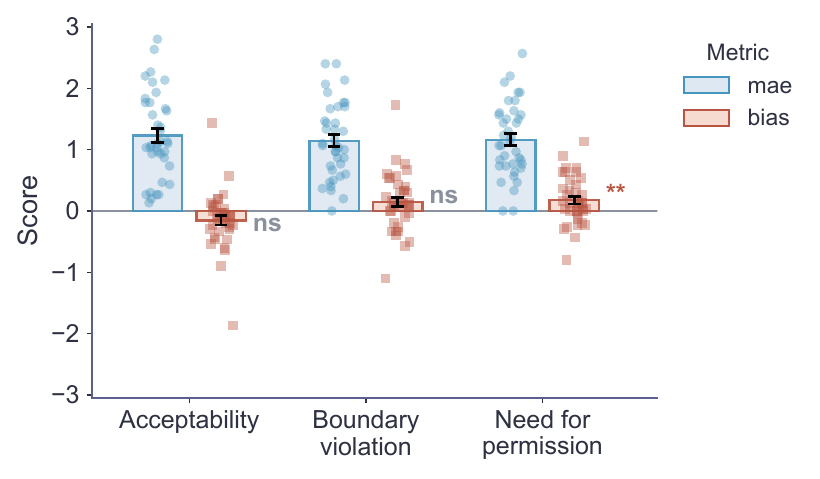}
    \caption{Human data co-owner alignment with owner judgments. MAE summarizes misalignment. Signed bias is prediction minus owner response. Significance markers for signed bias are based on unadjusted one-sample tests; Holm-adjusted $p$ values are reported in the text.}
    \Description{A grouped plot comparing human-partner alignment with owner judgments on three outcomes: acceptability, boundary violation, and need for permission. For each outcome, light bars and points show mean absolute error (around 1.1 to 1.3) and darker bars and points show signed bias (near zero). Individual dyad points are overlaid. Unadjusted significance annotations mark need for permission as significant, while the other two outcomes are labelled non-significant.}
    \label{fig:human_overall}
\end{figure*}

\subsection{Effects of Contextual Factors}

\textbf{User-related factors yielded limited and mixed associations with acceptability misalignment.} An exploratory room-level Welch test revealed that rooms with high information-specific familiarity showed significantly lower human acceptability misalignment than rooms with low information-specific familiarity (MAE Acc$= 1.004$ vs. $1.456$; median split at $6.05$; Welch $t(35.27)=-2.09$, unadjusted $p= .044$). However, dyad relationship types showed no significant effect on acceptability alignment error, whether evaluated via a comparison between friends and romantic partners (friend $M= 1.353$ vs. romantic partner $M= 1.107$; Welch $t(31.62)= 1.09$, $p= .285$) or via a multivariable trial-level model ($\beta = -0.109$, 95\% CI $[-0.308,0.090]$, $p= .283$; Figure~\ref{fig:coowner_prediction}).

\begin{figure*}[!htbp]
    \centering
    \subfloat[Familiarity to data owner.]{
        \includegraphics[width= 0.33\textwidth]{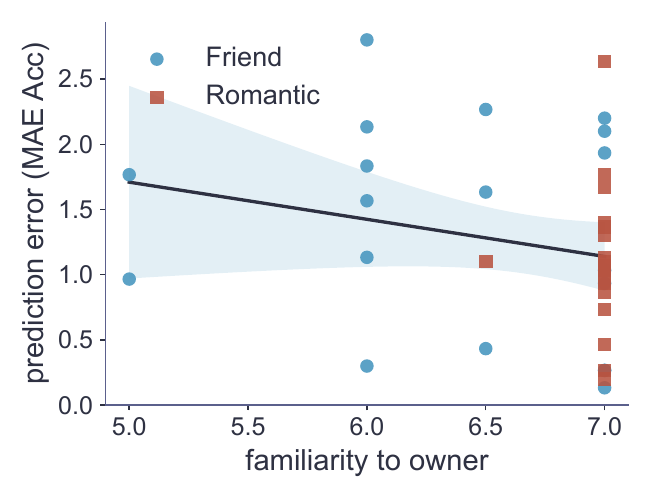}
    }
    \subfloat[Familiarity to data type.]{
        \includegraphics[width= 0.33\textwidth]{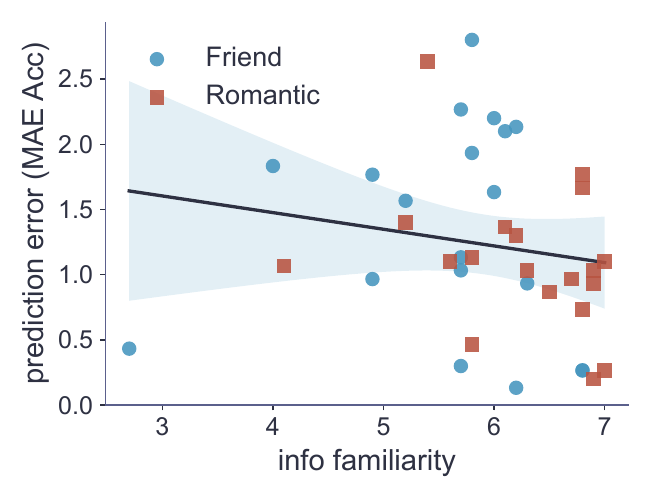}
    }
    \subfloat[EQ.]{
        \includegraphics[width= 0.33\textwidth]{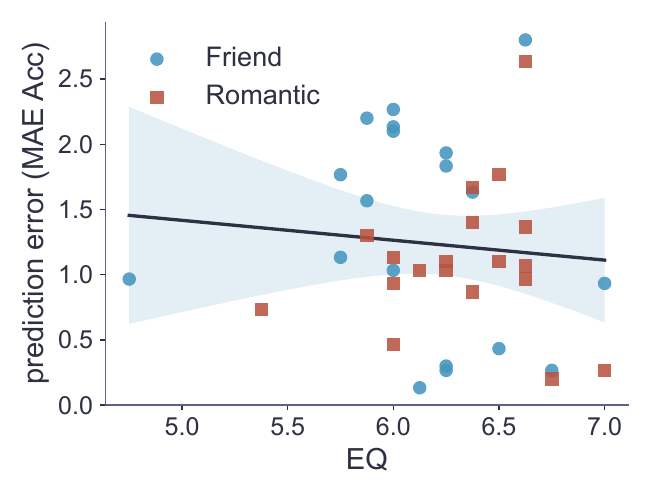}
    }
    \caption{The correlation of (a) familiarity to data owner, (b) familiarity to data type, and (c) EQ with misalignment (MAE). The points denote dyad-mean scores, the lines denote regression lines, and the shaded area denotes 95\% CI.}
    \Description{Three scatterplots relating predictors to misalignment. Panel (a) plots interpersonal familiarity against error; panel (b) information-type familiarity against error; panel (c) EQ score against error. Points distinguish friend dyads from romantic dyads, with a fitted trend line and shaded uncertainty band in each panel. Trends are mostly weak or modest, showing limited association between these traits and alignment error.}
    \label{fig:correlation_person}
\end{figure*}

\textbf{However, individual user-related traits such as interpersonal familiarity and the exploratory EQ score did not show significant dyad-level correlations with alignment.} As shown in Figure~\ref{fig:correlation_person}, general familiarity with the data owner showed no significant correlation with alignment error ($r= -.225$, $p= .175$; Figure~\ref{fig:correlation_person}a). Similarly, familiarity with specific information was not significantly correlated with alignment error ($r= -.170$, $p= .309$; Figure~\ref{fig:correlation_person}b). Furthermore, the exploratory EQ score showed no significant dyad-level correlation with alignment ($r= -.095$, $p= .571$; Figure~\ref{fig:correlation_person}c).

\textbf{Among contextual factors, data category significantly influenced alignment, while transmission recipient was not significantly associated with alignment.} A Kruskal--Wallis test indicated that data types significantly influenced mean absolute error averaged across the three judgments (Figure~\ref{fig:transmission_recipient}; $H(17)=44.27$, $p<.001$), though post-hoc comparisons were non-significant. Items yielding the lowest descriptive mean error included \textit{SPM-coparticipant} (MAE$= 0.893$) and \textit{SPM-romantic} (MAE$= 0.826$), while \textit{beliefs/thoughts} (MAE$= 1.648$) and \textit{name} (MAE$= 1.611$) yielded the highest errors. In contrast, transmission recipients showed no significant main effect on alignment between data owners and co-owners (Friedman $\chi^2(2) = 4.53$, $p = .104$, with post-hoc comparisons remaining non-significant across all pairs (stranger vs. family $p = .179$; stranger vs. friend $p = .173$; friend vs. family $p = .971$). Descriptively, scenarios with stranger recipients had slightly lower MAE ($M = 1.06$) than family ($M = 1.24$) and friend ($M = 1.24$) scenarios. One potential explanation is that co-owners may rely on shared, intuitive heuristics for specific information and data recipient, but struggle for others where personal boundaries vary widely.

\subsection{Co-owner Alignment Regression}\label{sec:predict_align}

We further used regression analysis to more rigorously analyze what factors are correlated with alignment errors. 

\textbf{Multivariable regression identified familiarity with the owner as the primary participant-level correlate of co-owner acceptability misalignment.} To determine whether alignment operates as an individualized model or a model based on broader social norm, we modeled alignment accuracy using both participant-level and data-level factors. As depicted in Figure~\ref{fig:coowner_prediction}, most contextual variables showed no significant correlation with alignment errors, including data categories such as \textit{State \& Behavioral} data ($\beta = +0.133$, 95\% CI [-0.101, 0.368], $p = .265$) and \textit{Cognitive \& Intimate} data ($\beta = -0.097$, [-0.285, 0.091], $p = .311$). Similarly, recipient types (friends vs. strangers: $\beta = 0.050$, [-0.155, 0.255], $p = .633$; family vs. strangers: $\beta = 0.063$, [-0.142, 0.268], $p = .546$) and dyadic relationships (romantic vs. friend: $\beta = -0.109$, [-0.308, 0.090], $p = .283$) did not significantly affect alignment. Crucially, familiarity with the owner emerged as the only statistically significant predictor of co-owner acceptability misalignment ($\beta = -0.140$, 95\% CI $[-0.253,-0.028]$, $p= .015$), with greater familiarity associated with lower misalignment. In contrast, information-specific familiarity ($\beta = 0.024$, 95\% CI $[-0.063,0.111]$, $p= .589$) and the exploratory EQ score ($\beta = 0.009$, 95\% CI $[-0.087,0.105]$, $p= .850$) were not significant.

\section{RQ3: Modeling Interdependent Privacy with AI}\label{sec:rq3}

We finally examine whether AI is capable of behaving like a human data co-owner when deciding on third-party disclosure. We evaluate traditional ML models and LLMs across zero-shot, one-shot and LoRA-based fine-tuning settings. We focus on two targets: data owner acceptability, which determines agent behavioral appropriateness, and data co-owner acceptability, which determines an agent's ability to simulate and act on behalf of the co-owner. Traditional ML models represent rule-based methods for judging such disclosure, while LLMs with few-shot or fine-tuning settings are similar to strangers and ``familiar'' persons, respectively. This also tested whether prior decision data are required, and are enough for privacy-aware agents. 


\subsection{Effects of Models and Relational Context}


To understand how AI can assist interdependent privacy management, we examine (1) effects of models, comprising architectures and reasoning paradigms, and (2) effects of relational context. The former reflects an AI system's inherent capacity to simulate social perspective-taking and reason about relational boundaries, similar to human empathy. The latter reflects the influence of relational context, ranging from zero-shot prompts to fine-tuning, which parallels a human co-owner's familiarity with the target data owner. 

\subsubsection{Effects of Models}


\textbf{Machine learning models perform well on owner acceptability prediction, but less on co-owner misalignment prediction, potentially reflecting a lack of social relational reasoning.} Machine learning models based on structured features showed split capabilities across targets (Table ~\ref{tab:combined_performance}). When predicting owner's privacy boundaries, all ML models significantly outperformed the naive mean baseline ($p < .001$ across pairwise metrics). For acceptability (Table~\ref{tab:combined_performance}), \textit{linear regression} achieved the lowest across-subject MAE (1.369), \textit{GBDT} has the highest within-one accuracy (64.21\%), while \textit{random forest} led in exact match (22.46\%). Similar gains occurred for boundary violation (Table~\ref{tab:combined_performance}) and the need for consent (Table~\ref{tab:combined_performance}), and for non-across-subject conditions. 

\begin{table*}[!htbp]
\centering
\caption{Prediction performance for acceptability, boundary violation, and the need for consent under across and non-across evaluation.}
\label{tab:combined_performance}
\begin{tabular}{llcccccc}
\toprule
& & \multicolumn{3}{c}{Across} & \multicolumn{3}{c}{Non-across} \\
\cmidrule(lr){3-5}\cmidrule(lr){6-8}
Task & Model & MAE & Exact & Within $\pm1$ & MAE & Exact & Within $\pm1$ \\
\midrule
\multirow{4}{*}{\shortstack[l]{Acceptability}}
  & Mean baseline & 2.040 & 9.12\% & 26.75\% & 2.040 & 9.12\% & 26.75\% \\
  & Linear        & 1.369 & 21.84\% & 62.81\% & 1.356 & 23.16\% & 63.42\% \\
  & GBDT          & 1.380 & 20.26\% & 64.21\% & 1.341 & 21.49\% & 65.00\% \\
  & Random Forest & 1.451 & 22.46\% & 60.96\% & 1.318 & 27.63\% & 65.53\% \\
\midrule
\multirow{4}{*}{\shortstack[l]{Boundary\\violation}}
  & Mean baseline & 1.892 & 11.58\% & 43.25\% & 1.892 & 11.58\% & 43.25\% \\
  & Linear        & 1.307 & 21.14\% & 66.14\% & 1.286 & 24.12\% & 67.02\% \\
  & GBDT          & 1.319 & 19.21\% & 67.28\% & 1.257 & 21.67\% & 69.47\% \\
  & Random Forest & 1.342 & 23.77\% & 64.91\% & 1.223 & 27.89\% & 71.05\% \\
\midrule
\multirow{4}{*}{\shortstack[l]{Need for\\consent}}
  & Mean baseline & 1.634 & 12.28\% & 45.88\% & 1.634 & 12.28\% & 45.88\% \\
  & Linear        & 1.271 & 24.30\% & 67.11\% & 1.266 & 25.18\% & 67.19\% \\
  & GBDT          & 1.268 & 20.35\% & 70.53\% & 1.209 & 24.47\% & 72.72\% \\
  & Random Forest & 1.264 & 25.88\% & 68.25\% & 1.166 & 29.39\% & 71.75\% \\
\bottomrule
\end{tabular}
\end{table*}

\textbf{However, performance decreased when predicting the data co-owner's acceptability misalignment (Table~\ref{tab:coowner_error}).} In across-subject evaluations, no models significantly reduced error over the naive baseline. \textit{GBDT} achieved the same MAE as the baseline (1.148 vs. 1.148; Holm-adjusted $p = 1.00$). \textit{Linear regression} and \textit{random forest} performed slightly worse (MAE = $1.189$ and $1.188$; Holm-adjusted $p = .010$ and $p = .30$, respectively). Although \textit{linear regression} and \textit{GBDT} produced lower exact-match accuracies than the baseline, these decreases were not significant after Holm correction (Holm-adjusted $p = .059$ and $p = .086$). \textit{Random forest}'s exact match was essentially unchanged (27.98\% vs.\ 27.72\%). In contrast, within-one accuracy declined significantly for all three models after Holm correction (\textit{linear regression} 72.28\%, \textit{GBDT} 77.89\%, \textit{random forest} 71.14\%, baseline 81.58\%; Holm-adjusted $p < .001$ for all).
In the non-across-subject setting, \textit{GBDT} and \textit{random forest} achieved lower MAE (1.090 vs. 1.148 and 1.054 vs.\ 1.148; Holm-adjusted $p = .032$ for both), but within-one accuracy remained significantly below baseline for all models (\textit{linear regression} 74.04\%, \textit{GBDT} 78.07\%, \textit{random forest} 76.23\%, baseline 81.58\%; Holm-adjusted $p < .001$, $p = .001$, $p < .001$, respectively).
Together, these results show that conventional classifiers offer little improvement in predicting co-owner misalignment, particularly in across-subject settings. While attribute-based feature mapping can effectively capture normative privacy expectations consistent with CI~\cite{nissenbaum2004privacy}, anticipating such misalignment may not rely solely on static contextual attributes. It may require modeling interpersonal perspective tasking and Theory of Mind~\cite{frith2005theory,park2023generative}.

\begin{table*}[!htbp]
\centering
\caption{Prediction performance for the data co-owner's acceptability error.}
\label{tab:coowner_error}
\begin{tabular}{lcccccc}
\toprule
& \multicolumn{3}{c}{Across} & \multicolumn{3}{c}{Non-across} \\
\cmidrule(lr){2-4}\cmidrule(lr){5-7}
Model & MAE & Exact & Within $\pm1$ & MAE & Exact & Within $\pm1$ \\
\midrule
Mean baseline & 1.148 & 27.72\% & 81.58\% & 1.148 & 27.72\% & 81.58\% \\
Linear        & 1.189 & 25.26\% & 72.28\% & 1.150 & 26.40\% & 74.04\% \\
GBDT          & 1.148 & 25.53\% & 77.89\% & 1.090 & 27.02\% & 78.07\% \\
Random Forest & 1.188 & 27.98\% & 71.14\% & 1.054 & 33.86\% & 76.23\% \\
\bottomrule
\end{tabular}
\end{table*}

\begin{figure*}[!htbp]
    \centering 
    \includegraphics[width=0.9\textwidth]{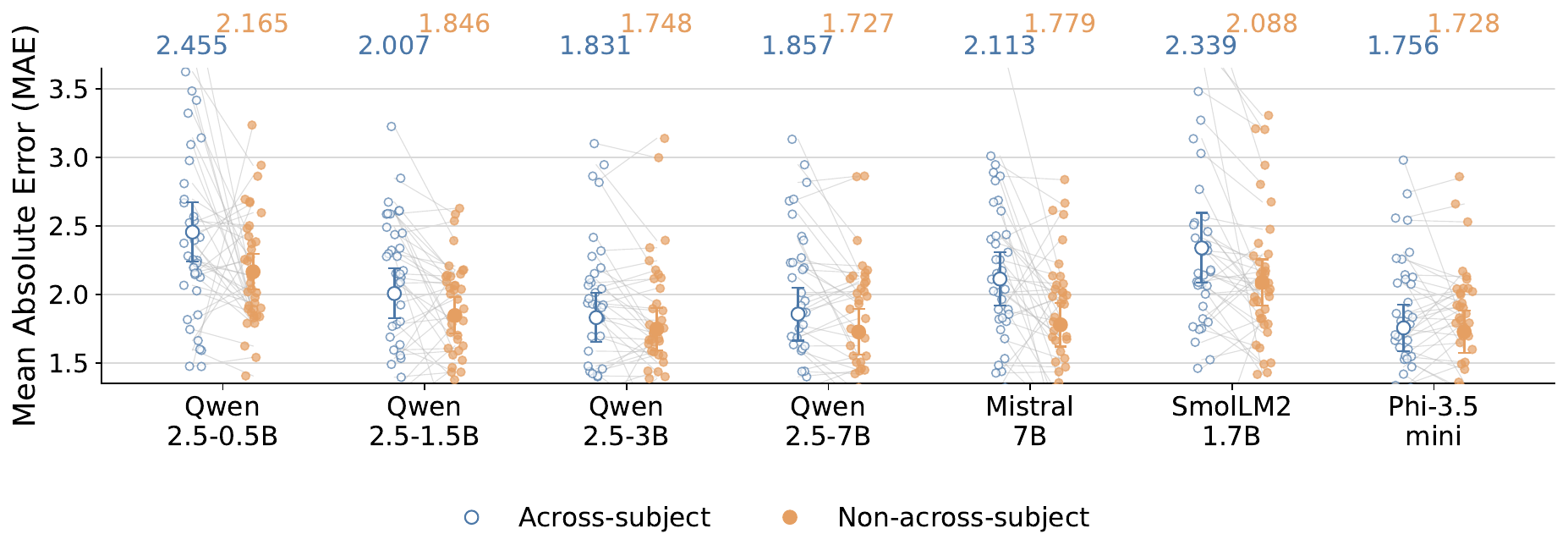}
    \caption{The prediction performance of LoRA models.}
    \Description{A plot of supervised fine-tuning (SFT) prediction accuracy for seven smaller models, comparing across-validation versus non-across validation. Each model has two markers: blue for across and orange for non-across. Non-across scores are higher for every model. Accuracies range from about 0.52 to about 0.70, with larger Qwen2.5 variants and Phi-3.5-mini among the stronger scores.}
    \label{fig:sft}
\end{figure*}

\textbf{Model scale affects relational privacy prediction up to a performance threshold.}
As in Figure~\ref{fig:sft}, model performance differed significantly in both the across-subject ($\chi^2_6=59.71$, $p<.001$) and the non-across-subject condition ($\chi^2_6=45.50$, $p<.001$). In the across-subject evaluation, \textit{Phi-3.5-mini} achieved the lowest MAE (1.756), followed by \textit{Qwen2.5-3B} (1.831) and \textit{Qwen2.5-7B} (1.857). In contrast, \textit{Qwen2.5-0.5B} and \textit{SmolLM2-1.7B} produced substantially larger errors (2.455 and 2.339, respectively). Pairwise comparisons showed that \textit{Qwen2.5-0.5B} had significantly higher MAE than \textit{Qwen2.5-3B} ($\Delta=0.624$, $p_{\mathrm{Holm}}<.001$), \textit{Qwen2.5-7B} ($\Delta=0.598$, $p_{\mathrm{Holm}}<.001$), and \textit{Phi-3.5-mini} ($\Delta=0.700$, $p_{\mathrm{Holm}}<.001$). \textit{SmolLM2-1.7B} also performed significantly worse than \textit{Qwen2.5-3B} ($\Delta=0.508$, $p_{\mathrm{Holm}}=.002$), \textit{Qwen2.5-7B} ($\Delta=0.482$, $p_{\mathrm{Holm}}=.009$), and \textit{Phi-3.5-mini} ($\Delta=0.584$, $p_{\mathrm{Holm}}<.001$). The non-across-subject evaluation yielded lower MAE for each model, with values ranging from 1.727 to 2.165. The lowest errors were observed for \textit{Qwen2.5-7B} (1.727), \textit{Phi-3.5-mini} (1.728), and \textit{Qwen2.5-3B} (1.748), whereas \textit{Qwen2.5-0.5B} remained the least accurate model (2.165). The reduction in MAE was model dependent ($\chi^2_6=18.76$, $p=.005$), indicating that access to contextual observations from the target subject did not provide a uniform improvement. Overall, these results suggest that relational privacy prediction benefits from models' capacity, but simply scaling parameter size yields diminishing benefits in aligning with human judgments.

\subsubsection{Effects of Relational Context}

We found that incorporating historical relational context yielded inconsistent gains in model performance across both owner acceptability and co-owner misalignment modeling.

\begin{figure*}[!htbp]
    \centering
    \includegraphics[width= 0.96\textwidth]{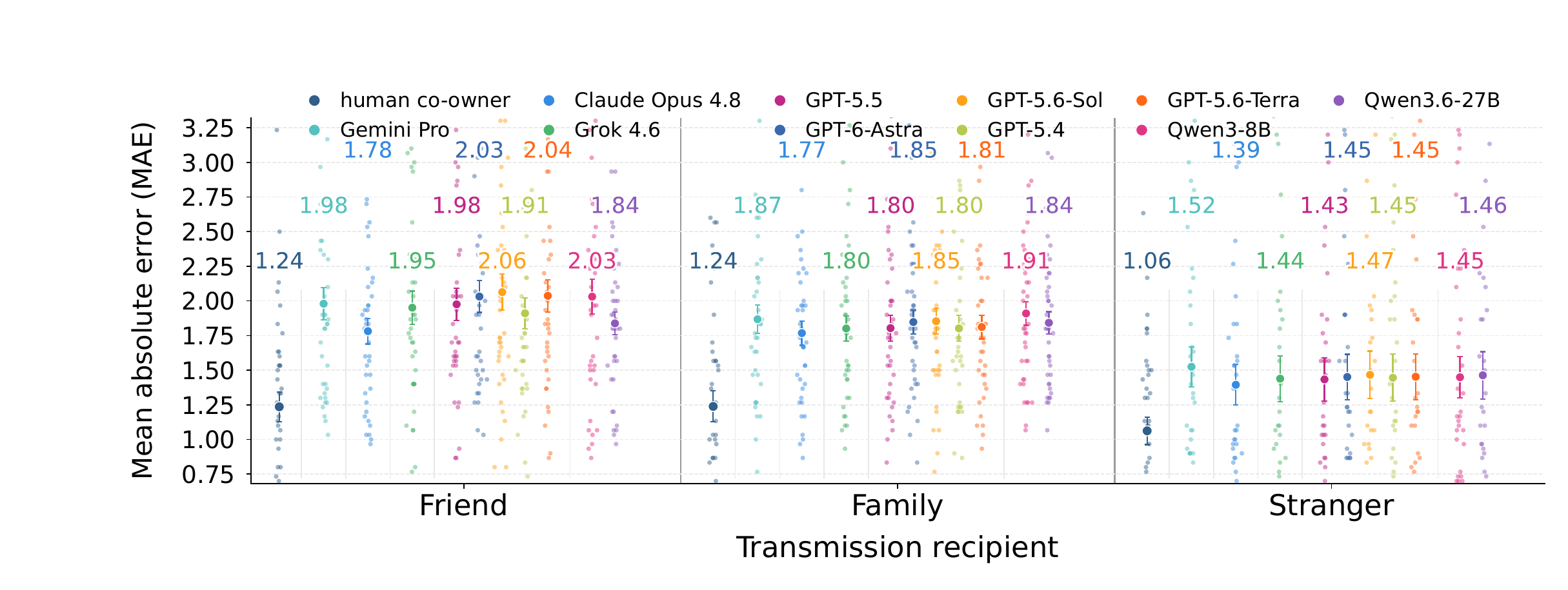}
    \caption{Zero-shot prediction error across transmission recipients for familiar human data co-owners and 10 LLMs. Small points show dyad-level mean absolute error (MAE), large markers show across-dyad means, and error bars indicate one standard error (SE). Lower MAE indicates predictions closer to the data owner's judgments.}
    \Description{A multi-panel plot of zero-shot mean absolute error for human data co-owners and ten large language models, split by disclosure recipient: friend, family, and stranger. Small points are dyad-level MAEs; large markers are means with standard-error bars. Human partners stay lowest, near about 1.1 to 1.25. All LLMs are higher, especially for friend and family recipients (roughly mid-1.7 to above 2.0), and somewhat better for stranger recipients (roughly mid-1.4 to mid-1.6). Lower MAE means closer to the owner's judgments. Points below the displayed y-axis range are clipped.}
    \label{fig:zeroshot_recipient}
\end{figure*}

\textbf{Zero-shot LLMs cannot model data owners' privacy expectations compared with human co-owners, revealing a deficit in capturing privacy heuristics.} Zero-shot prompting operationalizes an unacquainted agent lacking prior interaction history with the data owner. As shown in Table~\ref{tab:zeroshot_ai}, familiar human co-owners outperformed all 10 LLMs in MAE (1.179 vs. 1.648--1.796). This gap was mediated by relational closeness, as shown in Figure~\ref{fig:zeroshot_recipient}. LLMs performed poorly when evaluating disclosures to intimate relationships. Across models, average MAE was largest for disclosures to friends (1.960) and family (1.831), but dropped markedly for strangers (1.452). Conversely, human co-owners maintained stable performance across recipients (MAE= 1.236, 1.239, and 1.062 for friend, family and strangers). Similarly, LLM error was highest for acceptability (MAE= 1.946), intermediate for boundary violations (M= 1.727) and lowest for need for permission (1.569), while human error was uniformly distributed (MAE= 1.146-1.230). This suggests that AI models struggle for those judgments needing interpersonal contexts. For example, in close social relationships, privacy decisions may rely on implicit interpersonal heuristics that zero-shot LLMs cannot access.

\begin{table*}[!htbp]
\centering
\small
\caption{Zero-shot prediction performance of human co-owners and 10 LLMs. Lower MAE and higher accuracy indicate predictions closer to the data owner's judgments.}
\label{tab:zeroshot_ai}
\resizebox{\textwidth}{!}{%
\begin{tabular}{lccccccccccc}
\toprule
Metric & Human co-owner & Gemini Pro & Claude Opus 4.8 & Grok 4.6 & GPT-5.5 & GPT-6-Astra & GPT-5.6-Sol & GPT-5.4 & GPT-5.6-Terra & Qwen3-8B & Qwen3.6-27B \\
\midrule
MAE & 1.179& 1.791& 1.648& 1.730& 1.737& 1.777 & 1.794& 1.719& 1.767& 1.796 & 1.714\\
Exact Match & 40.4\%& 23.4\%& 22.6\%& 25.0\%& 24.4\%& 23.7\%& 24.4\%& 24.0\%& 24.0\%& 21.5\% & 23.2\%\\
Within $\pm 1$ & 70.7\%& 53.0\%& 53.6\%& 54.0\% & 53.7\%& 51.6\%& 52.0\%& 53.2\% & 52.4\% & 51.2\%& 52.2\%\\
\bottomrule
\end{tabular}%
}
\end{table*}

\begin{figure*}[!htbp]
    \centering
    \includegraphics[width=\textwidth]{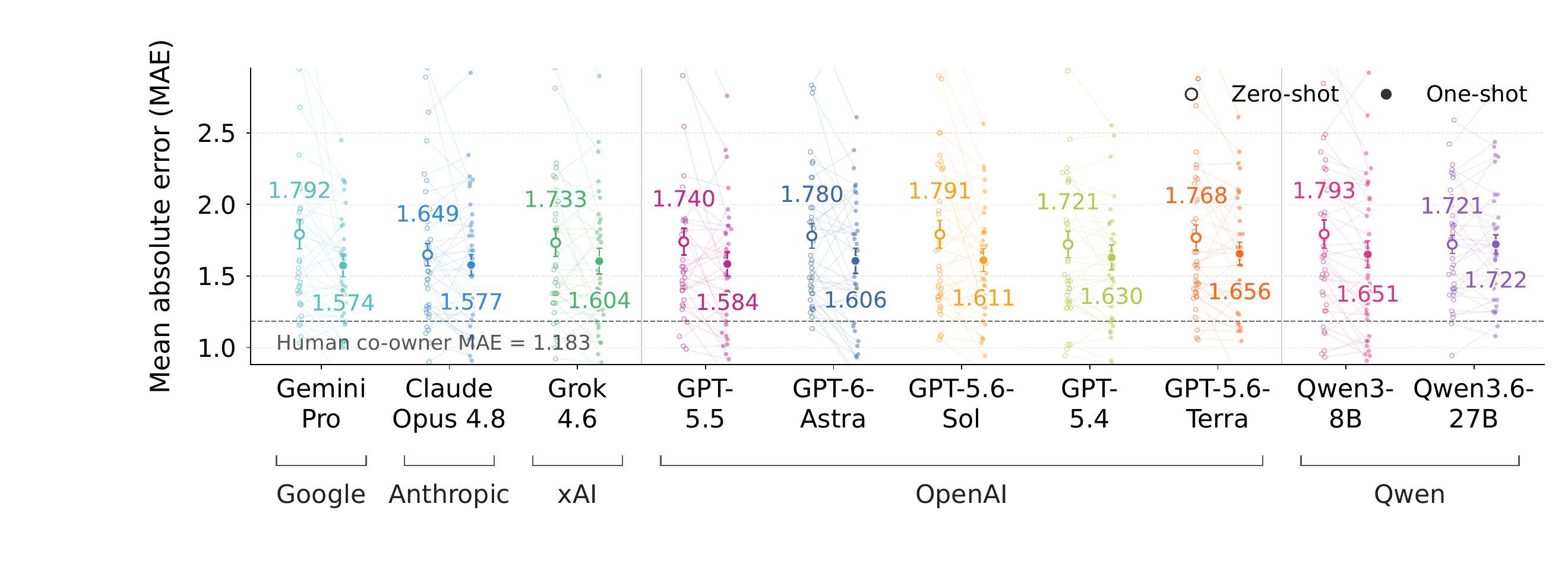}
    \caption{Zero-shot and one-shot prediction error across 10 LLMs. Small points show dyad-level MAE under zero-shot and one-shot prediction, with lines connecting matched observations from the same dyad within each model. Large markers show across-dyad means. Error bars indicate one standard error (SE). The dashed horizontal line denotes familiar human data co-owner performance. Lower MAE indicates better performance.}
    \Description{A model-by-model comparison of zero-shot versus one-shot prediction MAE for ten LLMs, with a dashed horizontal line for human-partner performance at MAE 1.20. For each model, light points are dyads, lines connect matched zero-shot and one-shot results, open markers show zero-shot means, and filled markers show one-shot means with error bars. One-shot usually lowers error; Gemini Pro and several GPT models improve clearly, while Qwen3.6-27B barely changes. All models remain above the human baseline.}
    \label{fig:oneshot_personalization}
\end{figure*}

\textbf{One-shot example is effective but limited in providing sufficient reasoning context for LLM.} As in Figure~\ref{fig:oneshot_personalization}, evaluation across 10 models showed that one-shot prompting reduced error ($\Delta$ MAE = -0.127, CI [-0.218, -0.042], t(37)=-2.779, p = .009). However, the benefits of the one-shot example were model-dependent (Friedman $\chi^2_9 = 20.23$, $p = .017$), where only \textit{Qwen3-8B} maintained significant improvement ($\Delta$ MAE = -0.143, $p = .019$). Conversely, models such as \textit{Qwen3.6-27B} showed negligible improvement (1.721 vs. 1.722, $p = .981$). Crucially, one-shot example remained insufficient to mitigate the gap between human and AI. Human co-owners achieved an MAE of 1.183, outperforming all one-shot prompting-based LLMs (MAE= 1.574--1.722, $p < .05$). 

\textbf{Contextual data used in LoRA fine-tuning showed limited and uneven generalization across unseen subjects.}
As shown in Figure~\ref{fig:sft}, the performance of non-across-subject setting compared with across-subject setting is model-dependent. This indicates that specific contextual data has limited effects in aligning with owner's privacy preferences (Friedman $\chi^2(6)=18.76$, $p=.005$). For example, improvement was significant for \textit{Qwen2.5-0.5B} ($\Delta$MAE $=0.290$, paired Wilcoxon $p=.020$), \textit{Mistral-7B} ($\Delta$MAE $=0.334$, $p=.002$), and \textit{SmolLM2-1.7B} ($\Delta$MAE $=0.251$, $p=.003$), but not for \textit{Qwen2.5-1.5B}, \textit{Qwen2.5-3B}, \textit{Qwen2.5-7B}, or \textit{Phi-3.5-mini} (all $p>.05$). Even under non-across-subject settings, LoRA models remained substantially less aligned than human co-owners. Their non-across-subject MAEs ranged from 1.727 to 2.165, compared with a human co-owner MAE of 1.183 ($p < .05$ for all models). This suggests that contextual data has limited effects in aligning with data owner's preferences.

Across unseen subjects, performance also varied with model size (Friedman $\chi^2(6)=59.71$, $p<.001$). \textit{Phi-3.5-mini}, \textit{Qwen2.5-3B}, and \textit{Qwen2.5-7B} obtained the lowest MAEs (1.756--1.857), while the smaller \textit{Qwen2.5-0.5B} and \textit{SmolLM2-1.7B} performed substantially worse (2.455 and 2.339, respectively, $p < .01$ compared with \textit{Qwen2.5-7B}, \textit{Qwen2.5-3B} and \textit{Phi-3.5-mini}). Indeed, all models did not differ with baselines predicting a constant number ($p > .05$). Finally, even the strongest across-subject model \textit{Phi-3.5-mini} (MAE $=1.756$) remained less accurate than human co-owners (human MAE = 1.183, $p < .05$). These results suggest that contextual data is insufficient for LLMs to learn relational privacy norms.

\subsection{Effects of Contextual Factors}
\label{sec:rq3-context}

We further examined whether LLM predictions and LLM--owner alignment varied across contextual factors. Using the matched zero-shot and one-shot data described in Section~\ref{sec:model-selection}, we fitted separate linear models for mean LLM acceptability predictions and mean absolute LLM--owner prediction error. Interpersonal familiarity, exploratory EQ, and dyad relationship type were modeled as separate fixed effects.

\textbf{LLMs reflected broad recipient and information-category patterns across prompting conditions.} LLMs rated disclosures to friends and family as more acceptable than disclosures to strangers (friend: $\beta=1.798$, 95\% CI $[1.689,1.907]$, $p<.001$; family: $\beta=2.837$, 95\% CI $[2.698,2.975]$, $p<.001$). These directions parallel the recipient differences in owners' judgments, although similarity in average coefficients does not imply prediction of an individual owner's response. LLMs also rated \textit{Cognitive \& Intimate} information as less acceptable than \textit{Identity \& Contextual} information ($\beta=-0.280$, 95\% CI $[-0.498,-0.061]$, $p=.013$), whereas \textit{State \& Behavioral} information did not differ significantly from \textit{Identity \& Contextual} information ($p=.925$). Information familiarity was positively associated with predicted acceptability ($\beta=0.126$, 95\% CI $[0.017,0.235]$, $p=.024$), while communication mode was not ($p=.446$).

\textbf{One-shot prompting shifted predictions upward and improved average alignment.} One-shot prompting increased predicted acceptability by $0.434$ points ($\beta=0.434$, 95\% CI $[0.186,0.681]$, $p=.001$) and reduced mean absolute acceptability error by $0.216$ points relative to zero-shot ($\beta=-0.216$, 95\% CI $[-0.334,-0.098]$, $p<.001$). No factor-level interaction remained significant after correction for either raw predictions (all $p_{\mathrm{Holm}}\geq.068$) or absolute error (all $p_{\mathrm{Holm}}\geq.111$). Therefore, the analyses support an overall improvement in alignment from one-shot prompting but do not establish that its effect differed across the measured contextual factors.

\begin{figure*}[!htbp]
  \centering
  \includegraphics[width=0.6\textwidth]{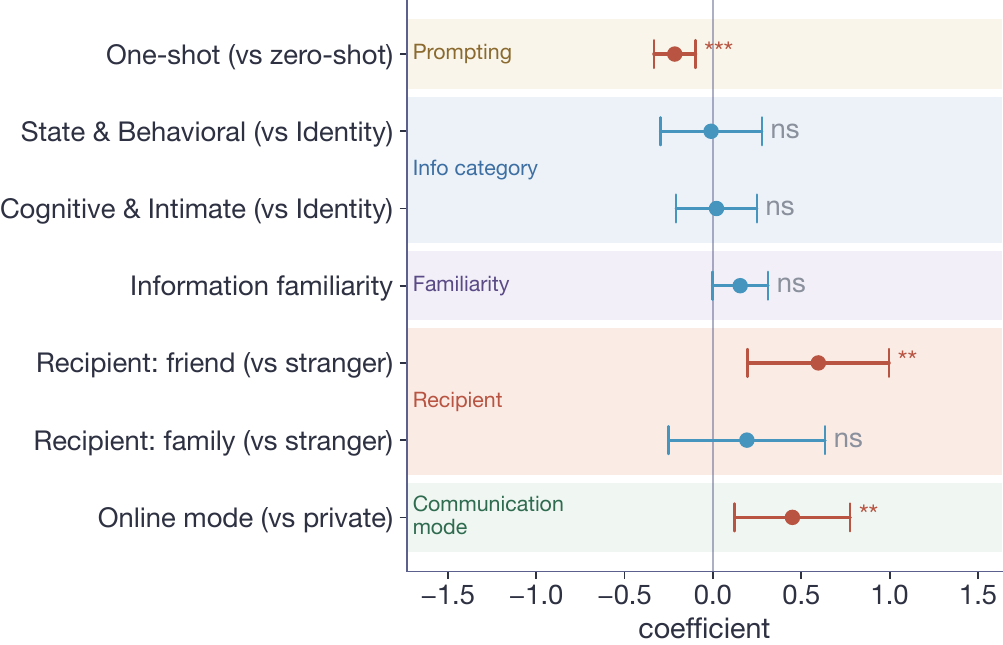}
  \caption{Regression of mean absolute LLM--owner acceptability error across matched zero-shot and one-shot conditions. Points show regression coefficients and error bars show 95\% confidence intervals. The model included information-owner fixed effects and dyad-clustered standard errors. Negative coefficients indicate lower absolute error relative to the corresponding reference or average condition.}
  \Description{A coefficient plot of mean absolute LLM--owner acceptability error across matched zero-shot and one-shot conditions. One-shot prompting is associated with lower error, whereas friend recipients and online communication mode are associated with higher error.}
  \label{fig:llm-contextual}
\end{figure*}

\textbf{Residual prediction error remained higher for friend recipients and online scenarios.} As shown in Figure~\ref{fig:llm-contextual}, absolute error averaged across prompting conditions was higher for friend than stranger recipients ($\beta=0.489$, 95\% CI $[0.123,0.856]$, $p=.010$). The family--stranger contrast was not significant ($\beta=0.180$, 95\% CI $[-0.235,0.594]$, $p=.385$). Online rather than private communication mode was associated with greater error ($\beta=0.443$, 95\% CI $[0.081,0.805]$, $p=.018$). Information category was not significantly associated with error (both contrasts $p>.40$), and the positive association with information familiarity are not significant either ($\beta=0.116$, 95\% CI $[-0.019,0.251]$, $p=.089$). Recipient also produced the strongest but non-significant interaction ($F(2,37)=4.65$, $p_{\mathrm{Holm}}=.111$). Overall, LLMs had similar broad population-level patterns in acceptability compared to humans while remaining less aligned with individual owners in scenarios with friends as recipients and online communications. 

\section{Discussions}

\subsection{Interdependent Privacy Boundaries}

We highlight unique contributions of our research to HCI and AI community both theoretically and empirically through contextualization. \textbf{\textit{First, our research advances the quantitative modeling of interdependent privacy acceptability, shifting the focus from isolated individual preferences to interpersonal data ownership.}} While prior studies modeled a single owner's privacy acceptability~\cite{zhang2026privacy,abdi2021privacy}, we augment them into a multi-party framework accounting for information shared by others. By integrating CI theory~\cite{nissenbaum2004privacy}, our findings show that interpersonal privacy boundaries are distinct constructs governed by relational dynamics and recipient contexts. 

\textbf{\textit{Second, we quantitatively augment CPM and privacy boundary theory~\cite{petronio2002boundaries}.}} Prior quantitative modeling primarily focused on the data owner's boundary definitions and measurements~\cite{guo2025not}. We extend this by modeling the acceptability towards data co-owner's transmission. For example, we show that familiar data co-owners often show a significant conservative bias for permission judgments than the data owners themselves, thereby generalizing personal privacy boundary management~\cite{guo2025not} to boundary management in interpersonal interactions.

\textbf{\textit{Third, we provide behavioral evidence that quantifies the misalignment between data owners and co-owners regarding the disclosure of shared data.}} While prior multi-stakeholder research highlights divergent privacy expectations between primary users and bystanders~\cite{wang2026mind,windl2022skewed}, our work sets an empirical basis for understanding interdependent privacy violations~\cite{humbert2019survey}. Specifically, greater familiarity with the owner was associated with lower co-owner acceptability misalignment, whereas information-specific familiarity and the exploratory EQ score were not statistically significant. This pattern suggests that relational knowledge of the owner may be relevant to anticipating the owner's acceptability judgments, although our observational analysis does not establish a causal effect. Furthermore, our findings reveal unawareness of interpersonal privacy norms, which distinguishes such risks from privacy violations rooted in accidental capture, as seen in bystander and multi-stakeholder contexts~\cite{bhardwaj2024focus,wang2026mind,abu2025they}, or malicious intent, as seen in intimate partner abuse literature~\cite{matthews2017stories,leitao2019anticipating}.

\textbf{\textit{Finally, we advance privacy preference prediction literature~\cite{yang2024feasibility} by expanding the predictive target to encompass co-owner's alignment gaps, and owner's expectations towards interdependent privacy.}} This yields implications for the design of privacy-aware AI agents. As LLMs increasingly mediate interpersonal communication, they require multi-stakeholder alignment frameworks. Our results indicate that models should go beyond aligning solely with the primary user~\cite{zhang2025towards} to dynamically align with secondary users~\cite{hussain2026idpbench,choksi2024groups}. By quantifying limitations of zero-shot AI predictions and the nuanced benefits of fine-tuning, we establish that future AI should integrate both broad social norms and owner-specific preferences to respect interdependent privacy boundaries.

\subsection{Generalizability}

We discuss our generalizability along three dimensions: cultural and demographic contexts, multi-stakeholder dynamics, and communication modalities.

First, privacy is culturally sensitive~\cite{xu2024dipa2,ur2013cross}. Our study sample was geographically diverse, comprising participants residing in South Africa (40.8\%), India (13.2\%), Egypt (10.5\%) and various Western countries, with only 50.0\% reporting English as their primary language. Besides, the underlying evaluation process of information flow based on sender, recipient and data type are largely similar across cultures, as evidenced by prior studies~\cite{zhang2026privacy,abdi2021privacy}. However, the exact thresholds of what is deemed acceptable for specific data types, such as financial or intimate information, may be tied to localized cultural norms, such as individualist and collectivist.

Second, our study focuses on a direct dyadic relationship and their disclosures to specific third-party recipients. The mechanisms of interdependent privacy and boundary coordination generalizes conceptually to extended social graphs. For instance, if a co-owner's friend learns the data, their disclosure to a stranger would still invoke similar principles. However, the predictive accuracy observed in our study cannot generalize to these multi-hop disclosures. For example, as information moves further away from the owner, the relational distance increases, and a bystander's perceived acceptability of disclosure will likely diverge significantly from a direct co-owner. 

Finally, our scenarios were grounded in online communication. The concepts regarding privacy boundaries can generalize to broader interactive modalities such as offline physical interactions. However, the performance of AI agents and human partners in predicting these boundaries is likely different. 

\subsection{Implications}

Our findings offer several theoretical, algorithmic, and interaction design implications for interdependent privacy.

\textbf{Theoretical: Computational modeling of interdependent privacy boundaries.} While CPM~\cite{petronio2002boundaries} conceptually characterizes boundary coordination, quantitative privacy research has primarily modeled isolated individual preferences~\cite{yang2024feasibility,zhang2025towards}. By synthesizing CPM with CI~\cite{nissenbaum2004privacy}, we show that interpersonal disclosure is governed by shared contextual norm. However, the substantial variance explained by individual factors highlights the limits of a purely normative model. Future research can build upon our results to develop standardized psychometric scales~\cite{marshall1974dimensions} and computational metrics that capture both normative consensus and boundary permeability.

\textbf{Theoretical: From single user to collective AI alignment.} Current personal assistants, such as Apple Intelligence, are optimized for single user agency, which execute tasks on behalf of the device owner. However, as assistants increasingly mediate interpersonal communication, third-party disclosure introduces interdependent privacy risks~\cite{humbert2019survey}. Aligning assistants exclusively with the operator risks violating the boundary of the data subject. Consequently, AI alignment frameworks should shift from aligning with the operator to collective alignment, which resolve multi-stakeholder tensions where the assistant acts as a bounded mediator balancing co-owner utility against owner vulnerability.

\textbf{Algorithmic: Global norms with local personalization.} Our evaluations in Section~\ref{sec:rq3} found that zero-shot LLMs and across-individual ML models generalize poorly to unseen owners, failing to close the human-AI gap even with SFT. This underscores that interpersonal privacy cannot be solved solely via learning from prior decisions. Instead, we advocate for a base model pretrained on context-dependent social norms, paired with lightweight personalization that calibrates to the specific owner's boundaries using both prior decisions and rich information such as chat histories. 

\textbf{Interaction: Calibrating interaction with asymmetric conservation.} Our findings show that familiar human co-owners consistently show asymmetric conservation, overestimating permission requirements and adopting stricter standards than owners themselves. AI assistants should inherit this human heuristic, defaulting to protective, consent-seeking actions when handling personal data, especially in ambiguous or high-intimacy contexts. To prevent notification fatigue, systems should employ risk-adaptive consent, sharing low-risk contextual data while keeping high-stakes data behind approval~\cite{zhang2024adanonymizer}.

\section{Limitations}

We acknowledge several limitations of this paper. First, our methodology relies on scenario-based survey, which is susceptible to the privacy paradox~\cite{kokolakis2017privacy}, wherein individual's stated privacy preferences diverge from their actual disclosure behaviors in real-world settings. We prioritize controlled contextual comparisons, and let participants disclose their data to increase ecological validity. Our study length in entries was similar to prior study~\cite{abdi2021privacy}, and we did not observe abnormal behavior in ratings (e.g., rating a constant number). Nevertheless, we still leave field tests for future work. 

Second, some trial-level analyses did not account for within-dyad dependence, which may affect significance interpretations. Nevertheless, we positioned these estimations as exploratory, and used these estimations to uncover trial-specific behavioral patterns that aggregated models might obscure. 

Third, the AI evaluations were not meant to be comprehensive. We only evaluated certain representative models with different parameter sizes, and simple fine-tuning methods. This is to provide initial insights into the prediction landscape. In a real-world deployment, integrated personal AI assistants may possess access to more private historical data, relational graphs and multimodal behavioral patterns. Our taxonomy also only tested 18 information types. Although they are widely used categories~\cite{zhou2025rescriber,guo2025not}, they obscures granular variations within each specific type, for example daily inquiries compared with medical diagnosis for health information. 

Finally, to serve an initial step, our dyadic samples were restricted in its relational scope, which primarily examined friend and romantic partners. There are many other relationships such as labmates and colleagues, which we defer to future work. 

\section{Ethical Considerations}

We acknowledge the potential ethical implications, and carefully considered these implications in accordance with the principles outlined in the Menlo Report~\cite{bailey2012menlo} and Belmont Report~\cite{beauchamp2008belmont}. Our studies all got the approval of our university's Institutional Review Board (IRB).

First, regarding \textbf{Respect for Persons}, we prioritized participant autonomy through transparent informed consent. Both members of each dyad were informed about the study's interdependent nature. To protect their data, participants were permitted and recommended to communicate specific private information outside the study interface. Any information shared privately in this manner was purposefully not recorded by our system. Besides, in the informed consent, we transparently disclosed to participants that we may use LLMs and machine learning models for processing their private data, and got their full acquaintance. 

Second, to ensure \textbf{Beneficence}, we minimized potential privacy risks and interpersonal harms. The study design avoided coercing the disclosure of highly sensitive personal data. All collected responses were anonymized and analyzed, with permissions only to the research team. 

Third, in adherence to \textbf{Justice}, we ensured equitable participant selection and fair compensation. Dyads were recruited globally via Prolific, ensuring diverse demographic representation without exploiting vulnerable populations. All participants were compensated fairly according to Prolific's recommended compensation standards. 

Finally, addressing \textbf{Respect for Law and Public Interest}, our methodology was approved by our university's IRB, and fully complied with the terms of service of the recruitment platform. 

\section{Conclusion}

We investigated interdependent privacy boundaries by comparing how information owners, human co-owners and AI models evaluate unauthorized data disclosure. Through a dyadic study of 38 friend and romantic-partner pairs evaluating 18 information types across three recipient relationships, we found that owners' privacy judgments are highly contextual. While familiar human partners show moderate predictive alignment, they overestimate the need for permission and act more cautiously than the owners themselves. Notably, greater interpersonal familiarity was associated with lower misalignment between human co-owners’ and owners’ acceptability judgments. Furthermore, traditional machine learning models and zero-shot LLMs struggle to generalize across individuals, underperforming familiar human partners. Although one-shot personalization improves AI prediction slightly, a significant human-AI performance gap remains. These findings emphasize that developing privacy-aware AI assistants requires systems that integrate broad contextual social norms with owner-specific relational preferences.




\clearpage
\bibliographystyle{ACM-Reference-Format}
\bibliography{sample-base}

\appendix

\section{Generative AI Usage}

We used ChatGPT and Gemini-3.1-pro for polishing the text of this paper, including checking grammar. We also used ChatGPT for refining the figures. All authors take full responsibilities for all this paper's content.

\section{Participant Demographics}
\label{app:demographics}

Table~\ref{tab:demographics} showed the demographics of participants. 

\begin{table*}[!htbp]
\centering
\small
\caption{Demographics of participants in the analytic sample with complete pair demographics ($N_{\text{dyads}} = 38$; $N_{\text{people}} = 76$). Age: $M = 29.7$, $SD = 8.0$, range 21--73.}
\label{tab:demographics}
\begin{tabular}{@{}lrr@{\hspace{1.8em}}lrr@{}}
\toprule
\textbf{Characteristic} & $\boldsymbol{n}$ & \textbf{\%} & \textbf{Characteristic} & $\boldsymbol{n}$ & \textbf{\%} \\
\midrule
\textit{Role in dyad} &  &  & \textit{Country of residence} &  &  \\
\quad A & 38 & 50.0 & \quad South Africa & 31 & 40.8 \\
\quad B & 38 & 50.0 & \quad India & 10 & 13.2 \\
\textit{Dyad relationship} &  &  & \quad Egypt & 8 & 10.5 \\
\quad Romantic partner & 38 & 50.0 & \quad United Kingdom & 4 & 5.3 \\
\quad Friend & 38 & 50.0 & \quad Chile & 4 & 5.3 \\
\textit{Gender} &  &  & \quad Canada & 4 & 5.3 \\
\quad Female & 41 & 53.9 & \quad Portugal & 2 & 2.6 \\
\quad Male & 35 & 46.1 & \quad Spain & 2 & 2.6 \\
\textit{Age} &  &  & \quad Israel & 2 & 2.6 \\
\quad 18--24 & 18 & 23.7 & \quad Brazil & 2 & 2.6 \\
\quad 25--29 & 29 & 38.2 & \quad Poland & 2 & 2.6 \\
\quad 30--34 & 16 & 21.1 & \quad Italy & 2 & 2.6 \\
\quad 35--39 & 6 & 7.9 & \quad Other & 3 & 3.9 \\
\quad 40+ & 7 & 9.2 & \textit{Primary language} &  &  \\
\textit{Ethnicity} &  &  & \quad English & 38 & 50.0 \\
\quad Black & 29 & 38.2 & \quad Arabic & 7 & 9.2 \\
\quad White & 20 & 26.3 & \quad Portuguese & 6 & 7.9 \\
\quad Mixed & 12 & 15.8 & \quad Spanish & 6 & 7.9 \\
\quad Asian & 10 & 13.2 & \quad Tamil & 4 & 5.3 \\
\quad Other & 4 & 5.3 & \quad Urdu & 2 & 2.6 \\
\quad Prefer not to say & 1 & 1.3 & \quad Bengali & 2 & 2.6 \\
\textit{Student status} &  &  & \quad Polish & 2 & 2.6 \\
\quad No & 42 & 55.3 & \quad Other & 9 & 11.8 \\
\quad Yes & 30 & 39.5 & \textit{Employment status} &  &  \\
\quad Unavailable / expired & 4 & 5.3 & \quad Full-time & 43 & 56.6 \\
 &  &  & \quad Part-time & 14 & 18.4 \\
 &  &  & \quad Unemployed (job seeking) & 12 & 15.8 \\
 &  &  & \quad Other & 6 & 7.9 \\
 &  &  & \quad Starting new job soon & 1 & 1.3 \\
\bottomrule
\end{tabular}
\end{table*}

\section{Recruitment Materials}
\label{app:recruitment}

Participants were recruited through Prolific using separate study listings for friend dyads and romantic-partner dyads. The two recruitment descriptions were identical except for the required relationship between the two participants.

\subsection{Relationship Eligibility}

\textbf{Friend condition.}
Participants were told that they must complete the task with another registered Prolific participant who was their friend, but not a romantic partner or family member. The two participants were paired by entering a room code created in our study system.

\textbf{Romantic-partner condition.}
Participants were told that they must complete the task with another registered Prolific participant who was their romantic partner. The two participants were paired by entering a room code created in our study system.

\subsection{Recruitment Description}

The following information was presented to participants in both recruitment conditions:

\begin{quote}
This study investigates paired privacy judgments. In this task, you will act as a ``data co-owner'' and evaluate different scenarios involving the disclosure of your partner's information. Your partner (the ``data owner'') will also evaluate scenarios involving the disclosure of their own information. During the study, you will exchange roles and consider privacy situations from both perspectives.

The main questionnaire sections must be completed independently. Please do not discuss your answers with your partner during this part, as we are interested in understanding each person's individual privacy judgments.

Please note that there are no right or wrong answers. We are interested in understanding different perspectives on privacy decisions.

\textbf{ About payment:} Each dyad received 22 GBP in total (11 GBP per participant). 
\end{quote}

\section{Study Procedure and Survey Materials}
\label{app:survey}

This appendix provides additional participant-facing details of the dyadic study to support reproducibility. The main study procedure is summarized in Figure~\ref{fig:procedure}.

\subsection{Pre-study Questionnaire}

Before the main scenario task, participants reported their relationship with their study partner (friend or romantic partner). We also collected interpersonal familiarity, information familiarity, and EQ assessment. Interpersonal familiarity captured how well participants knew their partner overall, whereas information familiarity captured how familiar they were with the specific information about their partner used in the study.

\paragraph{Exploratory four-item empathy measure.}
\label{sec:appendix_d_empathy}
Items (7-point: 1 = Strongly disagree … 7 = Strongly agree):
(1) I can usually tell when the other participant feels uncomfortable.
(2) I can imagine how the other participant would feel in a privacy-related situation.
(3) I notice subtle changes in other people’s emotions.
(4) I try to consider another person’s perspective before sharing information about them.

Items were informed by the Empathy Quotient and related perspective-taking measures but adapted for this study; they are not the original EQ scale or scoring. Personal EQ score is the mean of the four positive items per person. Then we define dyad-level EQ score from the average of the two partners', which is used for analysis. We treat this measure as exploratory.

\subsection{Dyadic Task Instructions}

Participants completed the study in pairs. Within each dyad, one participant served as the \textit{information owner}, while the other served as the \textit{data co-owner}. The information owner evaluated around the data co-owner's disclosure of their own information to a third party. The data co-owner was asked to evaluate their disclosure of data owner's information.

Participants were instructed to complete the main questionnaire independently and not discuss their answers with their partner during the task.

When personal information needed to be communicated between partners, participants could either enter it through the study interface or share it privately outside the platform. Information shared privately was not recorded by the study system.

\subsection{Scenario Presentation}

Each dyad was assigned five randomly selected information types from the 18-type pool reported in Table~\ref{tab:information_types}. Each information type was presented with three possible third-party recipients: a stranger, a friend, and a family member.

A scenario therefore followed the general structure:

\begin{quote}
The data co-owner shares a specified type of the information owner's information with a specified third-party recipient without first obtaining the owner's permission.
\end{quote}

The sender, information owner, communication channel, and absence of prior permission were held constant across scenarios. Information type and recipient relationship varied across trials.

Each ownership direction contained $5$ information types $\times$ $3$ recipients $=15$ scenario evaluations. Participants then exchanged roles and completed another 15 evaluations, resulting in 30 scenario evaluations per dyad. Scenario order was randomized within each assigned set.

\subsection{Information Owner Questions}

For each scenario, the information owner answered three questions on 7-point Likert scales towards the data co-owner's private information sharing:

\begin{itemize}
    \item \textbf{Acceptability:} How acceptable was the information-sharing behavior? Higher scores indicated greater perceived acceptability.

    \item \textbf{Privacy Boundary Violation:} To what extent did the disclosure violate the owner's privacy boundary? Higher scores indicated stronger perceived boundary violation.

    \item \textbf{Need for Permission:} To what extent should the sharer obtain permission before disclosing the information? Higher scores indicated a stronger perceived need for permission.
\end{itemize}

\subsection{Data Co-owner Judgment Questions}

The data co-owner evaluated the same scenario but was instructed to judge around disclosing data owner's private information, using the same three 7-point dimensions: acceptability, privacy boundary violation and need for permission. Each owner response could be directly matched with the corresponding co-owner response.

\subsection{Role Switch}

After the first 15 scenario evaluations, participants exchanged roles. The previous information owner became the data co-owner, and the previous data co-owner became the information owner. The same procedure was then repeated for the second ownership direction.

\section{Additional Baseline Comparisons}
~\label{sec:appendix_e}
\begin{figure*}[!htbp]
    \centering 
    \includegraphics[width=\textwidth]{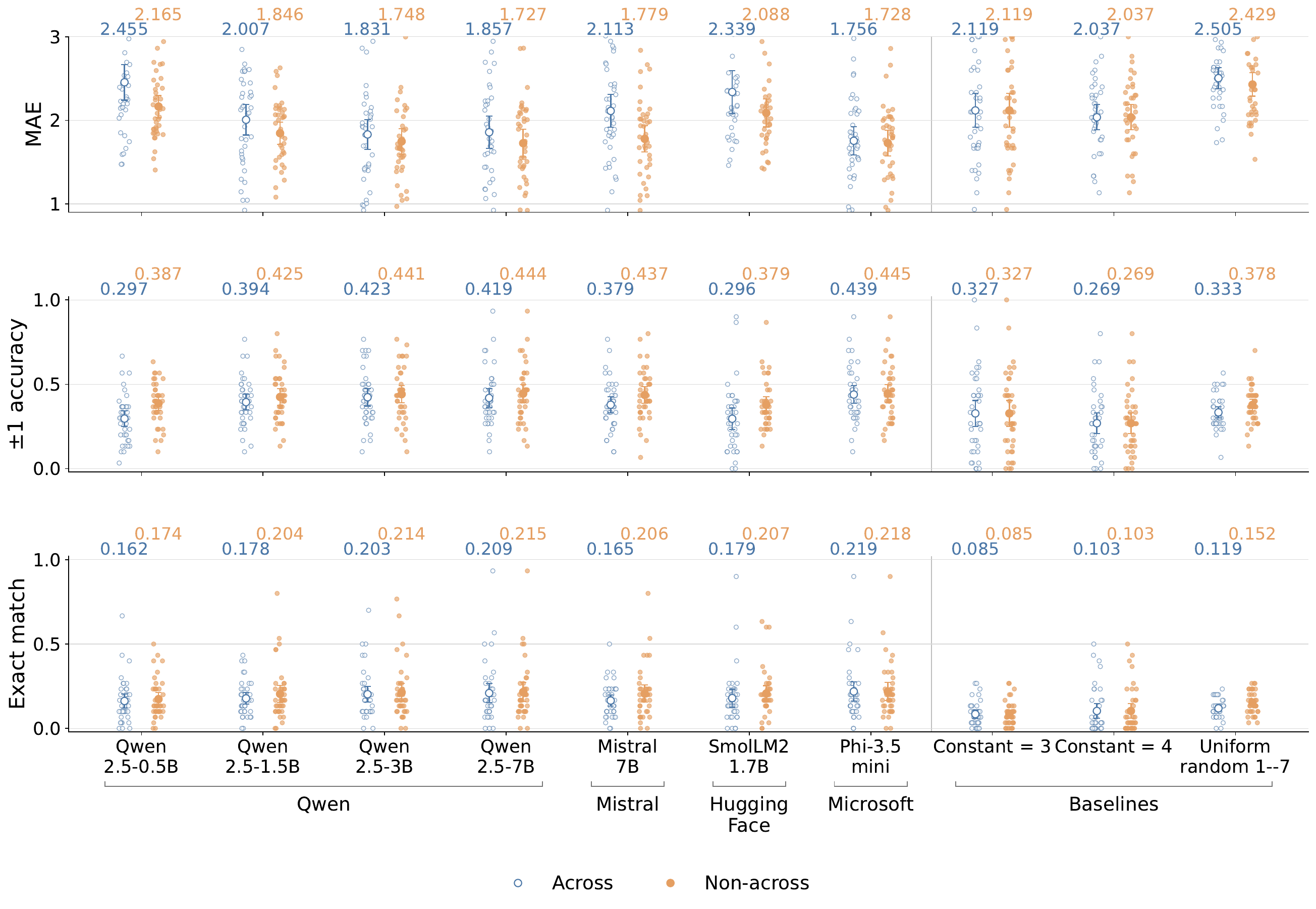}
    \caption{The performance of baseline models compared with LoRA-based fine-tuned models.}
    \Description{A three-panel figure stacked vertically that compares seven LoRA fine-tuned models with three simple baselines under across-subject and non-across-subject evaluation. The top panel shows mean absolute error, the middle panel within-one-point (±1) accuracy, and the bottom panel exact-match accuracy on the seven-point acceptability scale. Models are grouped by family: Qwen2.5-0.5B, 1.5B, 3B, and 7B; Mistral-7B; SmolLM2-1.7B (Hugging Face); and Phi-3.5-mini (Microsoft). Baselines are constant prediction of 3, constant prediction of 4, and uniform random integers from 1 to 7. Open blue markers denote across-subject means and filled orange markers denote non-across means, with small jittered points for individual observations and vertical whiskers for uncertainty. Constant-3 yields MAE 2.119, ±1 accuracy 0.327, and exact match 0.085; constant-4 yields MAE 2.039, ±1 accuracy 0.269, and exact match 0.091; uniform random is worst on MAE (about 2.43 to 2.51). Among fine-tuned models, stronger systems such as Phi-3.5-mini and larger Qwen2.5 variants reach MAE near 1.73 to 1.76 with higher ±1 and exact-match rates, while Qwen2.5-0.5B and SmolLM2-1.7B remain closer to—or worse than—the constant baselines on across-subject MAE (2.455 and 2.339).}
    \label{fig:lora}
\end{figure*}

To contextualize the performance of the AI models, we compared them with simple baselines such as predicting a fixed number on the original 1-7 rating scale, as shown in Figure~\ref{fig:lora}. Specifically, we evaluated constant predictions of 3 and 4 and a uniform random predictor that sampled an integer score from 1 to 7. All methods were evaluated using the same out-of-fold prediction setting under across-subject and non-across-subject validation settings. We report the same set of outcomes: mean absolute error (MAE), $\pm1$ accuracy, and exact-match accuracy. MAE measures the average absolute deviation between the predicted and observed owner ratings. $\pm1$ accuracy is the proportion of predictions within one Likert point of the observed rating, while exact match is the proportion of predictions that exactly matched the observed rating after rounding to the nearest integer.

The simple baselines provided substantially lower performance than the predictive models. A constant prediction of 3 yielded an MAE of 2.170, exact-match accuracy of 8.2\%, and within-one-point accuracy of 30.7\%. A constant prediction of 4 yielded an MAE of 2.039, exact-match accuracy of 9.1\%, and within-one-point accuracy of 26.8\%.

\end{document}